\documentclass[referee]{aa} 

\usepackage{amsmath, amssymb}

\usepackage{times}

\usepackage[dvips]{graphicx}
\usepackage{epsfig}
\usepackage{txfonts}
\usepackage{natbib}

\renewcommand{\div}{{\rm div}\,}

\begin{document}

\title{Non-linear evolution of the diocotron instability in a pulsar
  electrosphere: 2D PIC simulations.}

\author{J\'er\^ome P\'etri \inst{1}}

\offprints{J. P\'etri}

\institute{Observatoire Astronomique de Strasbourg, 11, rue de
  l'Universit\'e, 67000 Strasbourg, France.}

\date{Received / Accepted}

\titlerunning{Non-linear evolution of the diocotron instability}

\authorrunning{P\'etri}

\abstract
{(abridged) The physics of the pulsar magnetosphere near the neutron star surface
  remains poorly constrained by observations. Indeed, little is known
  about their emission mechanism, from radio to high-energy X-ray and
  gamma-rays. Nevertheless it is believed that large vacuum gaps exist
  in this magnetosphere, and a non-neutral plasma partially fills the
  neutron star surroundings to form an electrosphere in differential
  rotation.}
{According to several of our previous works, the equatorial disk in
  this electrosphere is diocotron and magnetron unstable, at least in
  the linear regime. To better assess the long term evolution of these
  instabilities, we study the behavior of the non-neutral plasma with
  help on particle simulations.}
{We designed a 2D electrostatic PIC code in cylindrical coordinates,
  solving Poisson equation for the electric potential. In the
  diocotron regime, the equation of motion for particles obeys the
  electric drift approximation. As in the linear study, the plasma is
  confined between two conducting walls. Moreover, in order to
  simulate a pair cascade in the gaps, we add a source term feeding
  the plasma with charged particles having the same sign as those
  already present in the electrosphere.}
{First we checked our code by looking for the linear development of
  the diocotron instability in the same regime as the one used in our
  previous work, for a plasma annulus and for a typical electrosphere
  with differential rotation.  To very good accuracy, we retrieve the
  same growth rates giving confidence for the correctness of our PIC
  code. Next, we consider the long term non-linear evolution of the
  diocotron instability. We found that particles tend to attract
  together to form small vortex of high charge density rotating around
  the axis of the cylinder with only little radial excursion of the
  particles.  This grouping of particles generates new low density or
  even vacuum gaps in the plasma column. Finally, in more general
  initial configurations, we show that particle injection into the
  plasma can drastically increase the diffusion of particles across
  the magnetic field lines.  Also, it has to be noted that the newly
  formed vacuum gaps cannot be replenished by simply invoking the
  diocotron instability.}


\maketitle

\section{INTRODUCTION}

The detailed structure of charge distribution and electric-current
circulation in the closed magnetosphere of a pulsar remains puzzling.
Although it is often assumed that the plasma fills the space entirely
and corotates with the neutron star, it is on the contrary very likely
that it only partly fills it, leaving large vacuum gaps between
plasma-filled regions. The existence of such gaps in aligned rotators
has been very clearly established by \cite{1985MNRAS.213P..43K,
  1985A&A...144...72K}.  Since then, a number of different numerical
approaches to the problem have confirmed their conclusions, including
some work by \cite{1989Ap&SS.158..297R}, \cite{1989Ap&SS.161..187S},
\cite{1993A&A...268..705Z}, \cite{1993A&A...274..319N},
\cite{1994ApJ...431..718T}, \cite{2002ASPC..271...81S}, and ourselves
\citep{2002A&A...384..414P}.  This conclusion about the existence of
vacuum gaps has been reached from a self-consistent solution of the
Maxwell equations in the case of the aligned rotator.  Moreover,
\cite{2001MNRAS.322..209S} have shown by numerical modelling that an
initially filled magnetosphere like the Goldreich-Julian model evolves
by opening up large gaps and stabilizes to the partially filled and
partially void solution found by \cite{1985MNRAS.213P..43K} and also
by \cite{2002A&A...384..414P}.  The status of models of the pulsar
magnetospheres, or electrospheres, has recently been critically
reviewed by \cite{2005RMxAC..23...27M}.  A solution with vacuum gaps
has the peculiar property that those parts of the magnetosphere that
are separated from the star's surface by a vacuum region are not
corotating and so suffer differential rotation, an essential
ingredient that will lead to non-neutral plasma instabilities in the
closed magnetosphere.
  
This kind of non-neutral plasma instability is well known to plasma
physicists (\cite{1980PhFl...23.2216O, Davidson1990,
  1992PhFlB...4.2720O}). Their good confinement properties makes them
a valuable tool for studying plasmas in laboratory, by using for
instance Penning traps.  In the magnetosphere of a pulsar, far from
the light cylinder and close to the neutron star surface, the
instability reduces to its non-relativistic and electrostatic form,
the diocotron instability.  The linear development of this instability
for a differentially rotating charged disk was studied by
\cite{2002A&A...387..520P}, in the thin disk limit, and by
\cite{2007A&A...469..843P, 2007A&A...464..135P} in the thick disk
limit. It both cases, the instability proceeds at a growth rate
comparable to the star's rotation rate.  The non-linear development of
this instability was studied by \cite{2003A&A...411..203P}, in the
framework of an infinitely thin disk model. They have shown that the
instability causes a cross-field transport of these charges in the
equatorial disk, evolving into a net out-flowing flux of charges.
\cite{2002ASPC..271...81S} have numerically studied the problem and
concluded that this charge transport tends to fill the gaps with
plasma. The appearance of a cross-field electric current as a result
of the diocotron instability has been observed by
\cite{2002AIPC..606..453P} in laboratory experiments in which charged
particles were continuously injected in the plasma column trapped in a
Malmberg-Penning configuration.
    
Numerical simulations of the non-linear development of the diocotron
instability have been investigated by different authors. For instance,
\cite{2003A&A...411..203P} used a numerical scheme similar to those
used for MHD codes.  This has the drawback to hardly handle vacuum
gaps which eventually are created in the electrosphere. To better deal
with these gaps, we decided to design a particle simulation code that
does not suffer from the presence of vacuum.  PIC methods have already
been successfull to investigate the diocotron instability
(\cite{1995AIPC..331..124N, 2003PhPl...10.3188Y}).  Problems with
injection have also been considered, see for instance
\cite{2007PhPl...14d2104B}.

In this paper, we extend the work initiated by
\cite{2007A&A...469..843P} and look for the non-linear evolution of
the diocotron instability by performing 2D electrostatic PIC
simulations.  The question on the influence of pair creation will also
be addressed by permitting injection of charged particles into the
system.  The paper is organized as follows.  In
Sect.~\ref{sec:Modele}, we briefly recall the initial setup of the
plasma column as done in \cite{2007A&A...464..135P}. In
Sect.~\ref{sec:ResultsWithout}, we show the results of our 2D PIC
simulations.  First we checked the code by computing some linear
growth rate for special cases like a constant density plasma column
for which we know exact analytical solutions, see for
instance~\cite{2007A&A...469..843P}.  Second, we check that we
retrieve the same linear growth rates for the electrosphere as those
found in \cite{2007A&A...464..135P}. Third, we let the system evolve
on long time scale and look for significant particle diffusion.
Finally, in a last set of runs, we injected some charged particles
into the system to study the behavior of the instability in the
presence of a source of charges, Sect.~\ref{sec:ResultsWith}. The
extreme case starting with vacuum will also be presented.  The
conclusions and the possible generalizations are presented in
Sect.~\ref{sec:Conclusion}.

\section{THE MODEL}
\label{sec:Modele}

Let us first remind the plasma configuration as described in
\cite{2007A&A...464..135P}. We study the motion of a non-neutral
plasma column of infinite axial extend along the $z$-axis. We adopt
cylindrical coordinates denoted by~$(r,\varphi,z)$ and define the
corresponding orthonormal basis vectors by~$(\vec{e}_{\rm
  r},\vec{e}_{\rm \varphi},\vec{e}_{\rm z})$. In the initial state
corresponding to an equilibrium, the plasma is only rotating in the
azimuthal direction, with no motion in the radial direction along~$r$
or vertical direction along~$z$.

\subsection{Plasma and field setup}

Unlike other magnetospheric models, our pulsar electrospheric plasma
is assumed to be completely charge-separated. So we consider a
single-species non-neutral fluid consisting of particles with
mass~$m_{\rm e}$ and charge~$q$ trapped between two cylindrically
conducting walls located at the radii $W_1$ and $W_2 > W_1$. The
plasma column itself is confined between the radii $R_1 \ge W_1$ and
$R_2 \le W_2$, with $R_1<R_2$.  This allows us to take into account
vacuum regions between the plasma and the conducting walls. Such
geometric configuration will also be useful to investigate particle
diffusion across magnetic field lines into this vacuum and will
clearly demonstrate the efficiency of the diocotron instability to
fill gaps with plasma as shown in some examples in the last section.

In order to simulate the presence of a charged neutron star generating
a radial electric quadrupole field coming from the rotating magnetic
dipole, the inner wall at~$W_1$ can carry a charge~$Q$ per unit length
such that its electric field is simply given by Maxwell-Gauss theorem,
namely, (we use MKSA units)
\begin{equation}
  \label{eq:E_r_ext}
  \vec{E}_\mathrm{w}(r) = \frac{Q}{2\,\pi\,\varepsilon_0\,r} \, 
  \vec{e}_\mathrm{r} .
\end{equation}
In the equilibrium configuration, the particle number density
is~$n(r)$ and the charge density is~$\rho_{\rm e}(r) = q \, n(r)$. In contrast
to earlier studies, the external magnetic field, along the~$z$-axis,
is not necessarily uniform but can decrease with radius
\begin{equation}
  \label{eq:Bz}
  \vec{B} = B_\mathrm{z}(r) \, \vec{e}_\mathrm{z} .
\end{equation}
The electric field is made of two parts, the first one arising from
the plasma column~$\vec{E}_\mathrm{p}$ itself, and the second one from
the inner conducting wall~$\vec{E}_\mathrm{w}$,
Eq.~(\ref{eq:E_r_ext}).  We assume that the electric field induced by
the plasma vanishes at $r=W_1$, i.e.  $\vec{E}_\mathrm{p}(W_1) =
\vec{0}$. Applying Maxwell-Gauss equation, we solve for the
equilibrium electric field and get
\begin{equation}
  \label{eq:Ep}
  \vec{E}_\mathrm{p}(r) = \frac{1}{\varepsilon_0\,r} \, \int_{W_1}^r 
  \rho_{\rm e}(r') \, r' \, dr' \, \vec{e}_\mathrm{r} .
\end{equation}
The total electric field, directed along the radial
direction~$\vec{e}_\mathrm{r}$ is therefore the sum
\begin{equation}
  \label{eq:ETot}
  \vec{E} = \vec{E}_\mathrm{p} + \vec{E}_\mathrm{w} = 
  E_\mathrm{r} \, \vec{e}_\mathrm{r} .
\end{equation}
In the non-relativistic diocotron regime, we do not solve the full set
of Maxwell equations but only the electrostatic part. This means that
the magnetic field perturbation induced by the plasma flow is
neglected.  Thus, the externally imposed magnetic field,
Eq.~(\ref{eq:Bz}), remains constant during the simulations.  This
assumption is justified when the plasma is confined well inside the
light-cylinder and when its rest mass energy density remains
negligible compared to the magnetic field energy density, in other
words
\begin{equation}
  \label{eq:Density}
  n \, m_{\rm e} \, c^2 \ll \frac{B^2}{2\,\mu_0}
\end{equation}

\subsection{Equation of motion}

The motion of the plasma is governed by the conservation of charge,
which in the non-neutral plasma case is equivalent to conservation of
the number of particles, the Maxwell-Poisson equation, and the
electric drift approximation, respectively
\begin{eqnarray}
  \label{eq:EqnMot1}
  \frac{\partial\rho_{\rm e}}{\partial t} + \div (\rho_{\rm e} \, \vec{v}) & = & 0 \\
  \Delta \phi + \frac{\rho_{\rm e}}{\varepsilon_0} & = & 0 \\
  \label{eq:Vit}
  \vec{v} & = & \frac{\vec{E} \wedge \vec{B}}{B^2} \\
  \label{eq:EqnMot4}
  \vec{E} & = & -\vec{\nabla} \phi .
\end{eqnarray}
$\phi$ is the usual electric potential associated with the total
electric field, Eq.~(\ref{eq:ETot}) (plasma distribution + externally
applied electric field).  We introduced the usual notation, $\rho_{\rm
  e}$ for the electric charge density, $\vec{v}$ for the velocity,
$(\vec{E}, \vec{B})$ for the electromagnetic field.  The set of
Eqs.~(\ref{eq:EqnMot1})-(\ref{eq:EqnMot4}) fully describes the
non-linear time evolution of the cold plasma in the non-relativistic
diocotron regime.  Note that particle inertia do not enter into the
above mentioned expressions. The gyromotion is averaged over the
fastest timescale, leading to the electric drift approximation, as
long as this drift speed remains smaller than the speed of light, in
other words, as long as $E<c\,B$. In the above mentioned equations
(\ref{eq:EqnMot1})-(\ref{eq:EqnMot4}), there is no reference to the
speed of light because we are in the non-relativistic regime.
Formally it corresponds to the limit $c\rightarrow+\infty$,
electromagnetic waves propagate instantaneously.  Therefore, at this
stage, it is impossible to check whether the electric field
intensity~$E$ remains smaller than $c\,B$ or not. The hypothesis
$E<c\,B$ is implicitly always true in our model because of the
electric drift approximation! Indeed $v<c$ in Eq.~(\ref{eq:Vit})
implies $E<c\,B$.  It is an implicit assumption that we presume to be
valid during the whole simulations.  To include relativistic effects,
we would have to solve the full set of Maxwell equations as was
already done to investigate the relativistic stabilization of the
diocotron instability in the linear regime, see for
instance~\cite{2007A&A...469..843P}. This would introduce a timescale
related to the speed of light~$c$.  Performing 2D relativistic PIC
simulations of the diocotron instability is the purpose of ongoing
work and will be exposed in a forthcoming paper.

\subsection{Code description}

We designed an algorithm using standard techniques for particle in
cell (PIC) codes as detailed for instance in two textbooks
\citep{Birdsall2005,Hockney1988}.  We use explicit schemes to advance
in time both the particles and the electric field. A description of
our own code on how to evolve field and particles is briefly exposed
in the following subsections.

\subsubsection{Integration of the equation of motion}

We assume that the inner and outer conductors act mechanically as
solid walls in such a way that particles are reflected at these
boundaries. The equation of motion for particles in the electric drift
approximation takes a very simple form allowing to solve immediately
for the velocity vector, see Eq.~(\ref{eq:Vit}). From a computational
point of view, this equation of motion, Eq.~(\ref{eq:Vit}), is
integrated with a leapfrog scheme, which is second order in time. More
precisely, denoting $t^n=n\,\Delta t$ the time at the $n$-th iteration
where $\Delta t$ is the time step, positions are computed at integer
times $t^n$ whereas velocities are computed at half-integer times
$t^{n+1/2}= (n+1/2)\,\Delta t$ for each particle labelled by the
index~$i$ such that
\begin{eqnarray}
  \label{eq:Numerics1}
  \vec{v}_i^{n+1/2} & = & \frac{\vec{E}_i^{n+1/2}
    \wedge \vec{B}_i^{n+1/2}}{{(B_i^{n+1/2})}^2} \\
  \label{eq:Numerics2}
  \frac{\vec{r}_i^{n+1}-\vec{r}_i^{n}}{\Delta t} & = & \vec{v}_i^{n+1/2}
\end{eqnarray}
The electromagnetic field at half-integer time is evaluate from the
position of the particles at the same time, positions updated
according to the speed known only at preceding times~$t^{n-1/2}$ like
\begin{equation}
  \label{eq:NumericsEB1}
  \vec{r}_i^{n+1/2} = \vec{r}_i^{n} + \frac{\Delta t}{2} \, \vec{v}_i^{n-1/2} 
\end{equation}
Then, the electromagnetic field at half-integer time following from
these positions at the location of particle number~$i$ are
\begin{eqnarray}
  \vec{E}_i^{n+1/2} & = & \vec{E}(r_i^{n+1/2}) \\
  \vec{B}_i^{n+1/2} & = & \vec{B}(r_i^{n+1/2})
\end{eqnarray}
We use a Lagrangian description which means that no grid is associated
with the position of the particle. Because the initial velocity and
position are given for $t=0$ by~$\vec{v}_i^{0}$ and~$\vec{r}_i^{0}$,
we initialize the speed of the particles by first doing half a time
step forwards in order to find
\begin{eqnarray}
  \label{eq:Initialisation}
  \vec{r}_i^{1/2} & = & \vec{r}_i^0 + \frac{\Delta t}{2} \,
  \vec{v}_i^0 \\ 
  \vec{v}_i^{1/2} & = & \frac{\vec{E}_i^{1/2} \wedge \vec{B}_i^{1/2}}{{(B_i^{1/2})}^2} 
\end{eqnarray}
We then evolve positions and velocities according to
Eqs.~(\ref{eq:Numerics1}),~(\ref{eq:Numerics2}). Note that in all the
simulations the electric and magnetic fields do not explicitly depend
on time.

\subsubsection{Field solver}

Poisson equation is solved on an Eulerian grid by Fourier transform in
the azimuthal direction~$\varphi$ and a finite difference scheme for
the radial dependence. Note that in Eq.~(\ref{eq:Numerics1}) both the
electric and the magnetic fields are evaluated at a half-integer
time~$t^{n+1/2}$. Because the electric field is derived from a
potential~$\phi$, we need only to consider a scalar Poisson equation.
The potential~$\phi$ as well as the charge density are developed into
real $\cos/\sin$ Fourier series as
\begin{eqnarray}
  \label{eq:Fourier}
  \phi(r,\varphi,t) & = & \sum_{m=0}^{N_\varphi/2-1} \phi_m^c(r,t) \, 
  \cos (m\,\varphi) + \phi_m^s(r,t) \, \sin( m\,\varphi) \\
  \rho_e(r,\varphi,t) & = & \sum_{m=0}^{N_\varphi/2-1} {\rho_e}_m^c(r,t) \, 
  \cos (m\,\varphi) + {\rho_e}_m^s(r,t) \, \sin (m\,\varphi)
\end{eqnarray}
$N_\varphi$ is the number of discretization points in the azimuthal
direction.  Thus Poisson equation transforms into a set of ordinary
differential equations for the unknown functions $\phi_m^{c/s}(r)$ as
follows
\begin{equation}
  \label{eq:FourierPoisson}
  \frac{1}{r} \, \frac{d}{dr} \left( r \, \frac{d\phi_m^{c/s}}{dr}
  \right) - \frac{m^2}{r^2} \, \phi_m^{c/s} =
  - \frac{{\rho_e}_m^{c/s}}{\varepsilon_0}
\end{equation}
Because of the linearity of the Poisson equation, the azimuthal modes
decouple into a set of independent ordinary differential equations for
each natural integer~$m$. Using a classical finite difference
discretization in radius, we derive a set of matrices for each couple
of unknown functions~$(\phi_m^c, \phi_m^s)$ to be solved by standard
linear algebra techniques.  However, the structure of the
discretization matrices is tridiagonal allowing to use numerical
algorithm that are more efficient, see for instance~\cite{Press2002}.
Moreover, as outer boundary condition, we insure a vanishing electric
potential at the outer wall which is expressed as $\phi_m^{c/s}(W_2) =
0$ whereas at the inner boundary, we enforce either a vanishing
derivative $d\phi_m^{c/s}(W_1)/dr = 0$ corresponding to nullifying the
normal component of the electric field for the vacuum or plasma annuli
cases, or the corotation electric field for the electrosphere, in
order to adjust smoothly to the revolution of the compact object
(assuming ideal MHD).

Charges are deposited on the Eulerian grid~$(r,\varphi)$ in order to
deduce the source terms~${\rho_e}_m^{c/s}$ for Poisson equation.  The
electromagnetic field is evaluated on particle positions by 2D linear
interpolation between points in this Eulerian grid~$(r,\varphi)$.

\subsubsection{Particle injection}
 
Finally, we simulate the pair creation mechanism in the magnetosphere
by injecting charged particles somewhere inside the system. The
deposition is uniform in the azimuthal direction and prescribed to be
at a given radius and rate, both fixed by the user.  Typically we took
an injection radius~$R_{\rm inj} = 2.0 \, W_1$.  Formally, the shape
function can be though as a delta-Dirac distribution in the radial
direction, let us say $S(r,\varphi) = 2\,\pi\,r\,\mathcal{F} \,
\delta(r - R_{\rm inj})$. $\mathcal{F}$ is the particle flux.
Nevertheless, this is a user defined source function and can be taken
to behave very differently, that can also evolve in space and time.
There is no particular restriction to its shape. As a starting point,
we took the most simple dependence, constant in time and $\varphi$ and
sharply concentrated in radius.  To be more realistic, pair creation
should be incorporate by taking into account different processes like
photon-photon disintegration or photon-(strong) magnetic field
interaction (which is formally the same process as the former, as the
magnetic field can be though as a photon field from a quantum
mechanical point of view). Monte Carlo simulations would be a valuable
tool for such a study, but this will not be included in this work.

\section{RESULTS WITHOUT PARTICLE INJECTION}
\label{sec:ResultsWithout}

First, we checked our algorithm in different configurations by picking
out the linear regime from our runs and compare the growth rates with
those predicted by the linear analysis.

In order to keep the particle noise as weak as possible to really see
the linear stage evolving on a few order of magnitudes, we need to
start with an initial state with minimal energy. This is achieved by
using a so-called "quiet start" such that space is regularly filled
with particles, see for instance~\cite{Birdsall2005}.

The resolution of the spatial grid on which the electric potential is
solved and charge is deposited is $N_r\times N_\varphi = 200 \times
256$, if not otherwise specified. On average, at the beginning of the
runs, there are 25 particles per cell giving a total number of
1,280,000.  Time is given in units of $1/\omega_B$ where
$\omega_B=q\,B/m_{\rm e}$ is the cyclotron frequency. We also tried
different resolutions and find no qualitative discrepancy. So we
adopted the canonical values aforementioned.

Note that for all the runs shown below, the charge $Q$ carried by the
inner conductor is set to zero, $Q=0$. The purpose of this work is not
to make a parametric study of the diocotron instability behavior but
rather to point out its main non-linear characteristics for a closed
(without particle injection or loss) and an open (with particle
injection or losses) system.  Moreover, in this and the subsequent
sections, the magnetic field is uniform and constant in the whole
plasma column. Let us now discuss the main results.

\subsection{Plasma annulus}
\label{sec:ColNRel}

For completeness, we briefly recall the main characteristics of the
configuration. The magnetic field is constant and uniform outside the
plasma annulus, namely $B_{\rm z} = B_0$, for $r>R_2$. We solve
Maxwell-Poisson equation in the space between the inner and the outer
wall, $R_1 < r < R_2$.  The particle number density and charge density
are constant in the whole plasma column such that
\begin{equation}
  \label{eq:RhoDiocRect}
  \rho_{\rm e}(r) = \left \lbrace
    \begin{array}{lcl}
      0   & , & W_1 \le r \le R_1 \\
      \rho_0 = n_0 \, q = {\rm const} & , & R_1 \le r \le R_2 \\
      0   & , & R_2 \le r \le W_2
    \end{array}
  \right.
\end{equation}
To initiate the instability we disturb the axisymmetric configuration
by overlapping a tiny density perturbation with relative magnitude
(i.e. compared to the background density) of the order $h=10^{-6}$. To
pick out a given mode, the perturbation is modulated with the
appropriate azimuthal pattern~$m$. To be more precise, the initial
axisymmetric equilibrium position of each particle $(r_0,\varphi_0)$
is perturbed according to
\begin{equation}
  \label{eq:Perturbation}
  \varphi_{\rm init} = \varphi_0 + h \, \cos(m\,\varphi_0)
\end{equation}

In this cylindrical geometry, an exact analytical solution of the
dispersion relation has been found for the diocotron instability, see
\cite{Davidson1990}.  We use these results to check our code, as was
already done in \cite{2007A&A...469..843P}.  Some exact analytical
eigenvalues are listed in Table~\ref{tab:DiocotronColonne} for
different radial plasma extensions and azimuthal modes.
\begin{table*}[htbp]
  \centering
  \begin{tabular}{cccc}
    \hline
    Mode $m$ & $d_1$ & $d_2$ & $\omega$ \\
    \hline
    2 & 0.4 & 0.5 & -3.772e-01 + 7.176e-02 \, i \\
    3 & 0.4 & 0.5 & -5.456e-01 + 2.267e-01 \, i \\
    4 & 0.4 & 0.5 & -7.216e-01 + 2.988e-01 \, i \\
    5 & 0.7 & 0.9 & -1.147e+00 + 5.787e-02 \, i \\
    7 & 0.6 & 0.7 & -9.315e-01 + 3.307e-01 \, i \\
    \hline
  \end{tabular}
  \caption{Eigenvalues~$\omega$ for the constant density 
    plasma annulus, for different modes~$m$ and different aspect ratios, 
    $d_1 = R_1 / W_2$, and $d_2 = R_2 / W_2$, obtained from 
    the analytical expression found in~\cite{Davidson1990}.}
  \label{tab:DiocotronColonne}  
\end{table*}
In order to quantify the growth rate of the diocotron instability in
our simulations, we estimate the total electrostatic energy in the
system by computing the integral
\begin{equation}
  \label{eq:Energie_Electrostatique}
  W_e(t) = \int_{W_1}^{W_2} \, \int_0^{2\pi} \varepsilon_0 \,
  \frac{E^2(t)}{2} \, r \, dr \, d\varphi  
\end{equation}
To clearly point out the linear growing phase, we calculate the
increase in electrostatic energy by seeking for the difference between
the actual and the initial electrostatic energy as
\begin{equation}
  \label{eq:DeltaWe}
  \Delta W_e(t) = W_e(t) - W_e(0)
\end{equation} 
Note that because $\Delta W_e(t)$ is a quadratic
expression with respect to the perturbed quantities, the linear growth
rate for the electrostatic energy increase will be twice the value
given in Table~\ref{tab:DiocotronColonne}.  

\begin{figure*}[htbp] 
  \centering
  \begin{tabular}{cc}
    \includegraphics[scale=0.7]{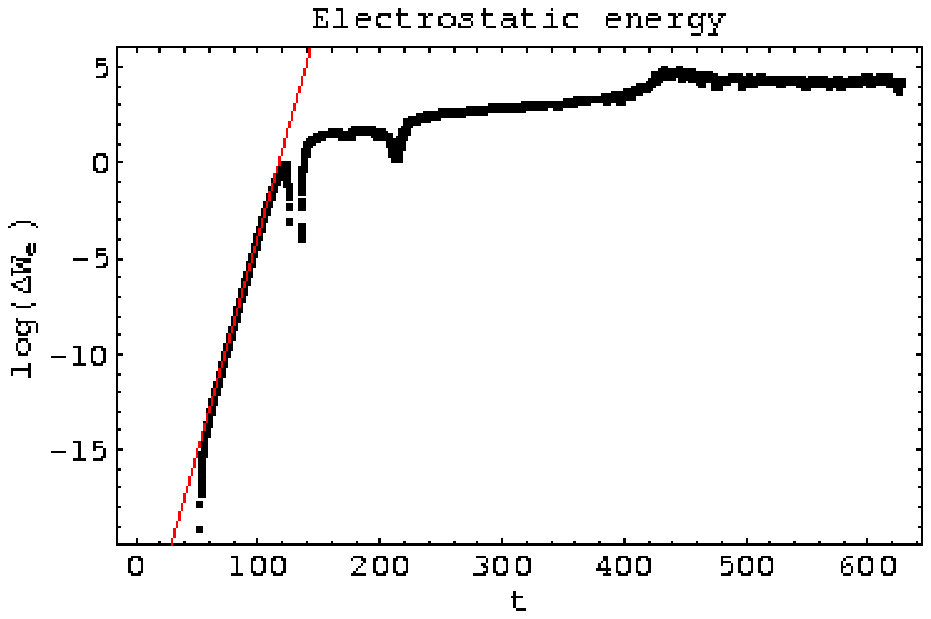} &
    \includegraphics[scale=0.7]{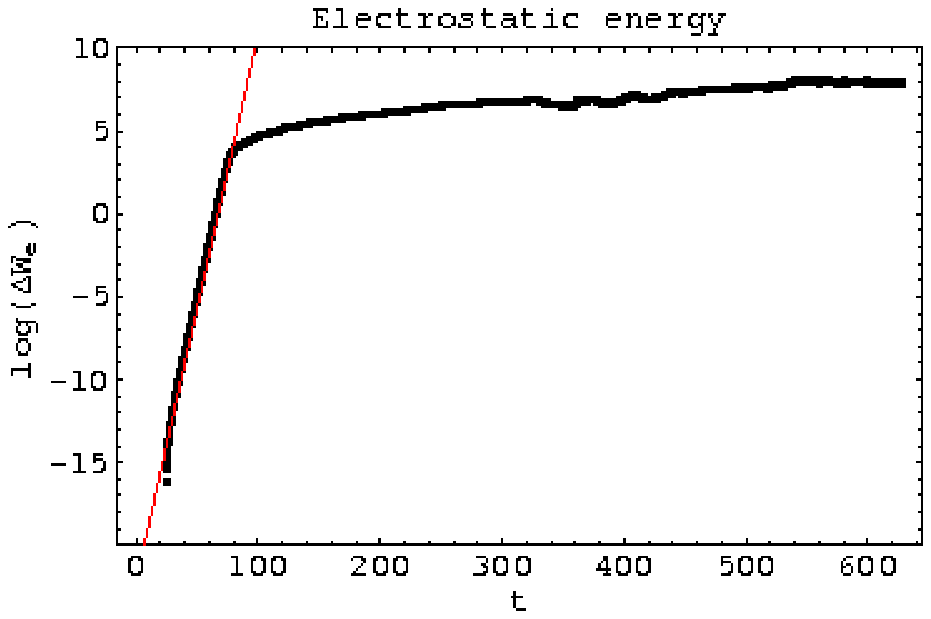} 
  \end{tabular}
  \caption{Evolution of the total electrostatic increase in
    energy~$\Delta W_e(t)$ in the constant density plasma column
    obtained from our PIC simulation (black dots) and compared with
    the linear growth rate of the diocotron instability obtained from
    Tab.~\ref{tab:DiocotronColonne} (red line), on the left for $m=3$,
    on the right for $m=7$. The linear relation holds for more than
    ten orders of magnitude.}
  \label{fig:Colonne_m37}
\end{figure*}

Two examples of run are now discussed. For the first run, the plasma
annulus is located between $R_1=4$ and $R_2=5$ with a perturbation
$m=3$, call it $A1$ while for the second run we took $R_1=6$, $R_2=7$
with a perturbation $m=7$, call it $A2$.  Let us have a look on the
evolution of the electrostatic energy given by Eq.~(\ref{eq:DeltaWe})
and depicted in Fig.~\ref{fig:Colonne_m37} on a logarithmic scale.

For the first run, at small times, let say $t<100$ and after a while,
the linear regime sets in and the measured growth rate is close to the
one predicted by linear analysis (red straight line), the perturbed
pattern grows at a speed predicted by the linear theory.  The plasma
column stays roughly confined between $R_1$ and $R_2$, no strong
density rearrangement is perceptible. Note also that no other mode is
excited, no cascade due to non-linear effects is observed because to
weak.

In some runs, as the first aforementioned, it can happen that the
initial electrostatic fluctuations decrease slightly before a
significant increase of many orders of magnitude, after a while,
inducing a~$\Delta W_e(t)<0$ at the beginning and thus making the
logarithm not defined at this early stage. In
Fig.~\ref{fig:Colonne_m37}, it is doubtful to give a physical
interpretation to this initial behaviour because it corresponds to
numerical round off error, close to the machine precision (we work
with double precision numbers, i.e. 16 significant digits).  Moreover,
the gaps can also be are probably due to the particle noise inherent
of the PIC method, proportional to the inverse of the square root of
the number of particles~$N$, (i.e. $\propto N^{-1/2}$). Tiny
fluctuations in the positions, especially particles receding to the
inner wall, tend to decrease the total electrostatic energy, at least
at the beginning of the simulations.

\begin{figure*}[htbp]
  \centering
  \begin{tabular}{cc}
    \includegraphics[scale=0.7]{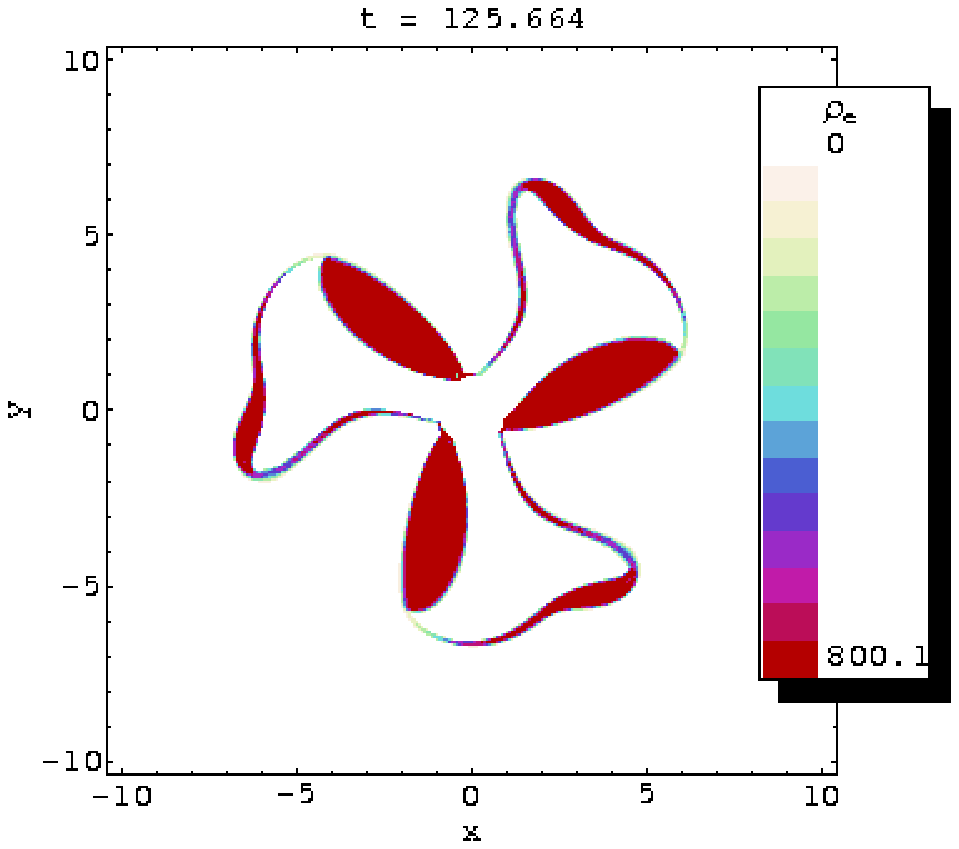} &
    \includegraphics[scale=0.7]{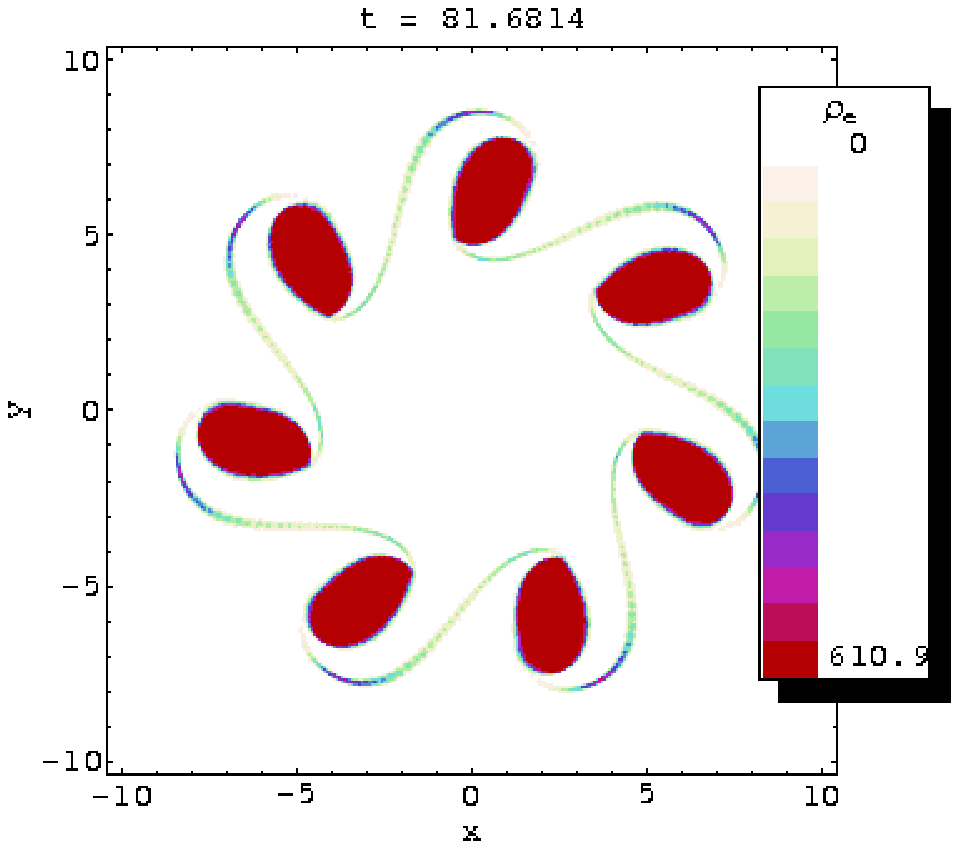} 
  \end{tabular}
  \caption{Snapshot of the charge density in the plasma column showing
    the $m=3$ pattern (on the left) and the $m=7$ pattern (on the
    right). The chosen time corresponds to the transition between the
    linear phase and the beginning of the non-linear regime,
    associated with the total electrostatic energy curves discussed in
    Fig.~\ref{fig:Colonne_m37}.}
  \label{fig:Colonne_m37_rho}
\end{figure*}

\begin{figure*}
  \centering
  \begin{tabular}{cc}
    \includegraphics[scale=0.7]{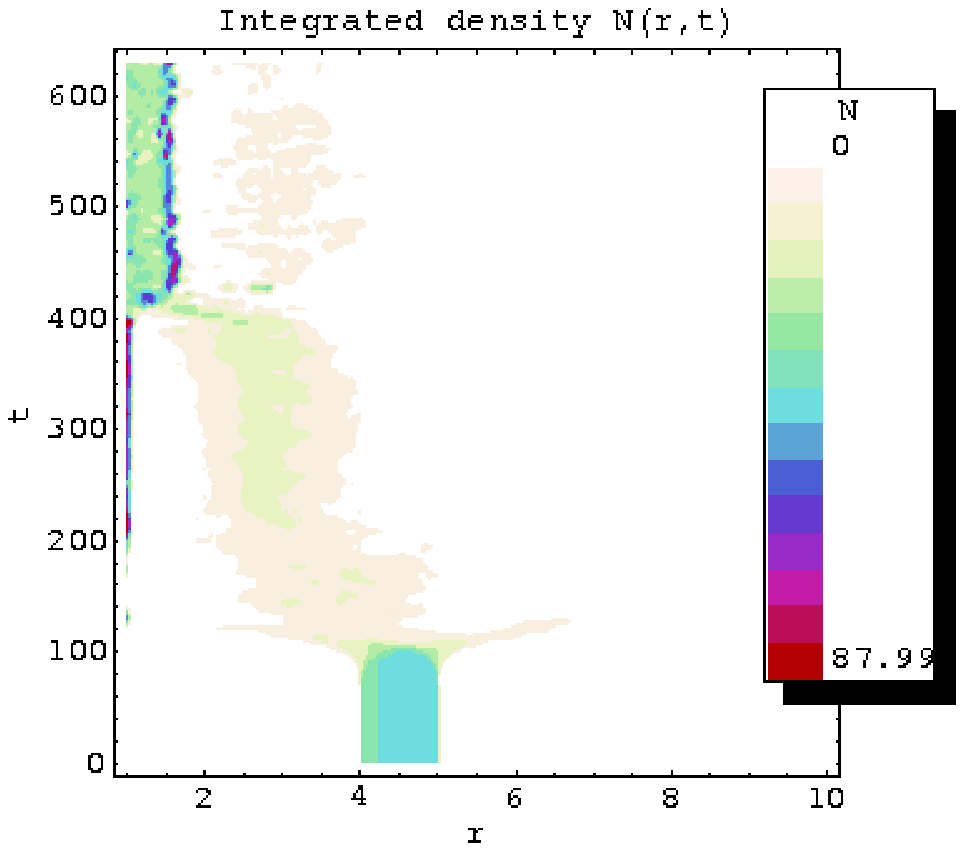} &
    \includegraphics[scale=0.7]{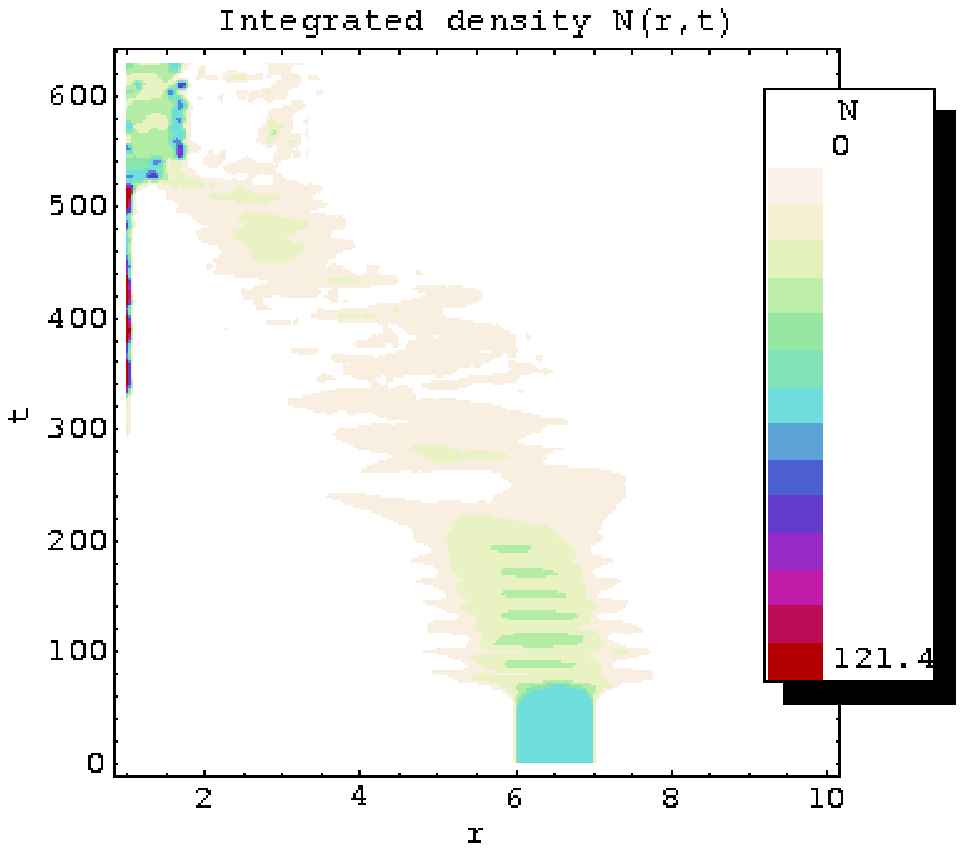} 
  \end{tabular}  
  \caption{Time evolution of the azimuthally integrated charge density
    $\mathcal{N}$ in the plasma column, on the left for $m=3$ and on
    the right for $m=7$. At the end of the runs, particles stay very
    close to the inner wall.}
  \label{fig:Flux_Densite_Colonne37}
\end{figure*}

In a second step, non-linear effects come into play and the initial
plasma distribution is destroyed. The azimuthal pattern with the
typical $m=3$ mode becomes clearly visible, left part of
Fig.~\ref{fig:Colonne_m37_rho}. Particles "merge" together to form
large vortices and leave behind them vacuum gaps which will not be
replenished. For some time, these vortices rotate around the axis of
the cylinder keeping their $m=3$ structure.  In a last step, almost
all particles hit the inner wall $W_1$, this happens when $t>400$. The
total electrostatic energy saturates, reaching a plateau at the latest
time.
When the non-linear effects set in, we are interested in the diffusion
of particle across magnetic field lines and on its consequences on a
longer timescale.  To get a better idea of this diffusion process, we
quantify it by computing the azimuthally integrated charge density as
\begin{equation}
  \label{eq:Densite_Charge}
  \mathcal{N}(r,t) = \int_0^{2\,\pi} \rho_{\rm e} \, r \, d\varphi
\end{equation}
It can be interpreted as an equivalent radial charge density. Indeed
\begin{equation}
  \label{eq:Charge_Int}
  \int_{W_1}^{W_2} \mathcal{N}(r,t) \, dr
\end{equation}
gives the total charge contained in the plasma column at time~$t$.
Some examples related to the simulations shown in
Fig.~\ref{fig:Colonne_m37} are given in
Fig.~\ref{fig:Flux_Densite_Colonne37}. We immediately see that for
$t<100$ (linear stage), the boundary of the plasma column remains
unchanged as discussed before.  In a second stage, non-linear effects
become important and strongly disturb the boundaries of the plasma
annulus.  Some very coherent patterns emerge while continuing to
rotate. During this period roughly $100<t<400$, the barycenter of the
charges moved closer to the inner wall as seen on the left part of
Fig.~\ref{fig:Flux_Densite_Colonne37}.  The structure implodes and
radial forces shift particles closer to the inner wall. In a final
stage, only few particles have moved outwards while the main fraction
drifted close to the inner boundary. Therefore we do not observe any
significant diffusion radially outwards.

The situation is very similar for the second run we show, namely A2.
Here also, the linear regime exists until $t\approx 60$ and the
observed growth rate is in close agreement with the linear analysis
(red straight line on the right part of Fig.~\ref{fig:Colonne_m37}).
The plasma column stays roughly confined between $R_1$ and $R_2$,
right part of Fig.~\ref{fig:Flux_Densite_Colonne37}, no strong density
rearrangement is perceptible during this linear phase. During the
transition from the linear to the non-linear state, the $m=7$
structure emerges in the density map as shown on the right part of
Fig.~\ref{fig:Colonne_m37_rho}.  Vacuum gaps are formed and seven
vortices rotate around the cylinder axis until some of them merge
together, destabilizing the whole system and shrinking down to the
inner wall.

These conclusions apply to all of our runs starting with an initial
equilibrium as a plasma annulus. Such behavior is expected because of
the confinement theorem which states that on average, the radial
excursion of particles is limited due to angular momentum conservation
of the system (particles + electromagnetic field),
see~\cite{Davidson1990}.

\subsection{Electrosphere}

Next, we switch to the results obtained for the electrosphere. As in
our previous works, the rotation profile is chosen to mimic the
rotation curve existing in the 3D electrosphere.  We remind that
different analytical expressions for the radial dependence of $\Omega$
are chosen by mainly varying the gradient in differential shear as
follows
\begin{equation}
  \label{eq:ProfilVit}
  \Omega(r) = \Omega_* \, ( 2 + \tanh[  \alpha \, ( r - r_0 ) ] \, e^{-\beta\,r^4} )
\end{equation}
\begin{table}[htbp]
  \centering
  \caption{Parameters for the three rotation profiles used
    to mimic the azimuthal velocity of the plasma 
    in the electrospheric disk.}
  \begin{tabular}{cccc}
    \hline
    $\Omega$ & $\alpha$ & $\beta$ & $r_0$ \\
    \hline
    $\Omega_1$ & 3.0 & $5\times10^{-5}$ & 6.0 \\
    $\Omega_2$ & 1.0 & $5\times10^{-5}$ & 6.0 \\
    $\Omega_3$ & 0.3 & $5\times10^{-5}$ & 10.0 \\
    \hline
  \end{tabular}
  \label{tab:Vitesse}
\end{table}
$\Omega_*$ is the neutron star spin and $r$ is normalized to the
neutron star radius, $R_*$. The values used are listed in
Table~\ref{tab:Vitesse}. In all cases, $\Omega$ starts from corotation
with the star $\Omega = \Omega_*$ at $r=1$, followed by a sharp
increase around $r=6$ for $\Omega_{1,2}$ and a less pronounced
gradient around $r=10$ for $\Omega_3$.  Finally the rotation rate
asymptotes twice the neutron star rotation speed for large radii, see
discussion and figures in~\cite{2007A&A...464..135P}.
\begin{table*}[htbp]
  \centering
  \begin{tabular}{ccc}
    \hline
    Model & Mode $m$ & $\omega$ \\
    \hline
    $\Omega_1$ &  3 & 6.669 + 1.8715 \, i \\
    $\Omega_1$ &  8 & 16.74 + 3.1925 \, i \\
    $\Omega_1$ & 15 & 30.98 + 1.6313 \, i \\
    $\Omega_2$ &  3 & 6.761 + 0.9759 \, i \\
    $\Omega_2$ &  5 & 10.90 + 0.765 \, i \\
    $\Omega_3$ &  3 & 5.643 + 0.3394 \, i \\
    \hline
  \end{tabular}
  \caption{Eigenvalues~$\omega$ for the electrospheric model $\Omega_1$,
    $\Omega_2$ and $\Omega_3$ for some azimuthal modes~$m$.}
  \label{tab:DiocotronElectrosphere}  
\end{table*}
\begin{figure*}[htbp]
  \centering
  \begin{tabular}{cc}
    \includegraphics[scale=0.7]{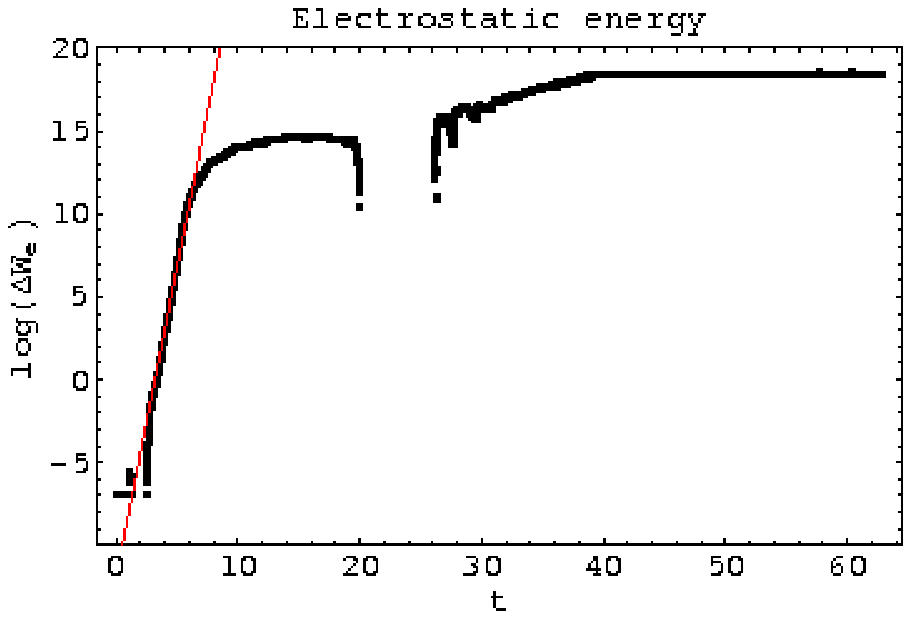} &
    \includegraphics[scale=0.7]{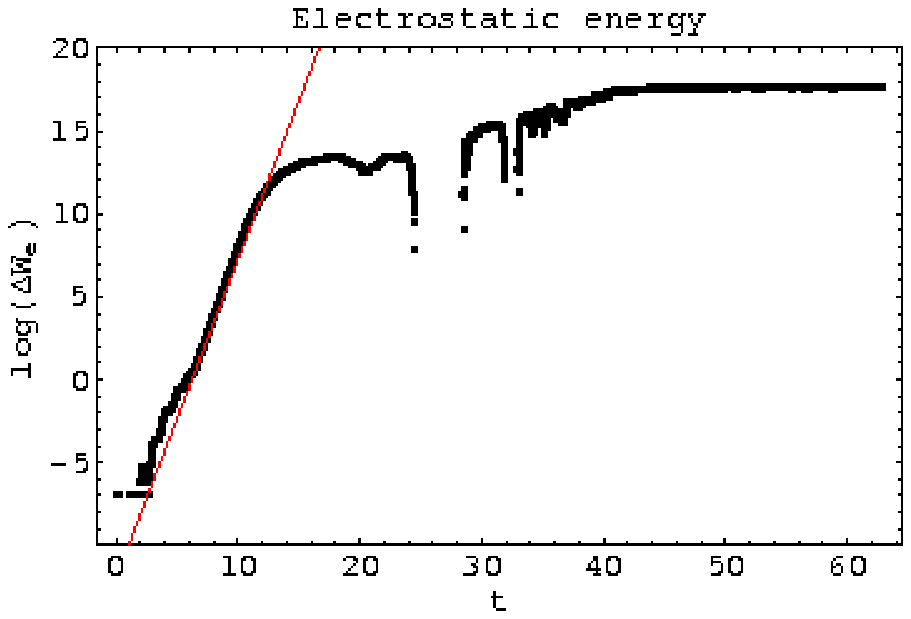} 
  \end{tabular}
  \caption{Evolution of the total electrostatic increase in energy for
    the differentially rotating plasma column obtained from our 2D PIC
    simulations (black dots) and compared with the linear growth rate
    obtained from Tab.~\ref{tab:DiocotronElectrosphere} (red line), on
    the left for the case $\Omega_1, m=3$ and on the right for
    $\Omega_2, m=3$.}
  \label{fig:Electrosphere_12}
\end{figure*}
\begin{figure*}[htbp]
  \centering
  \begin{tabular}{cc}
    \includegraphics[scale=0.7]{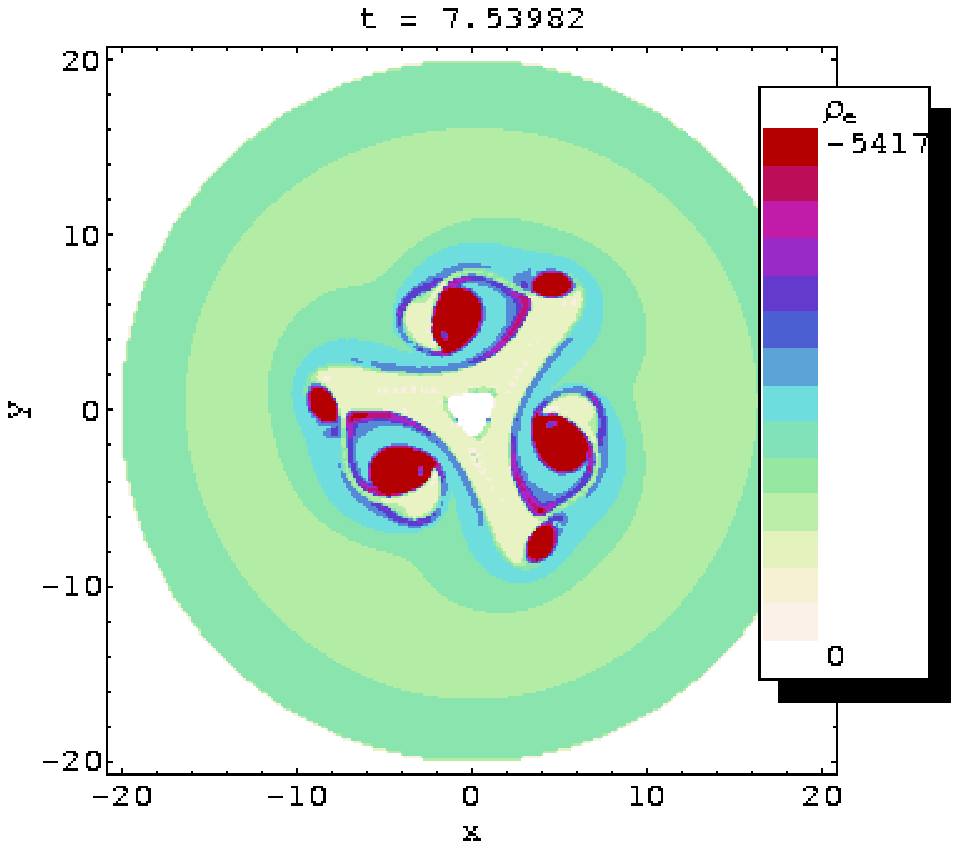} &
    \includegraphics[scale=0.7]{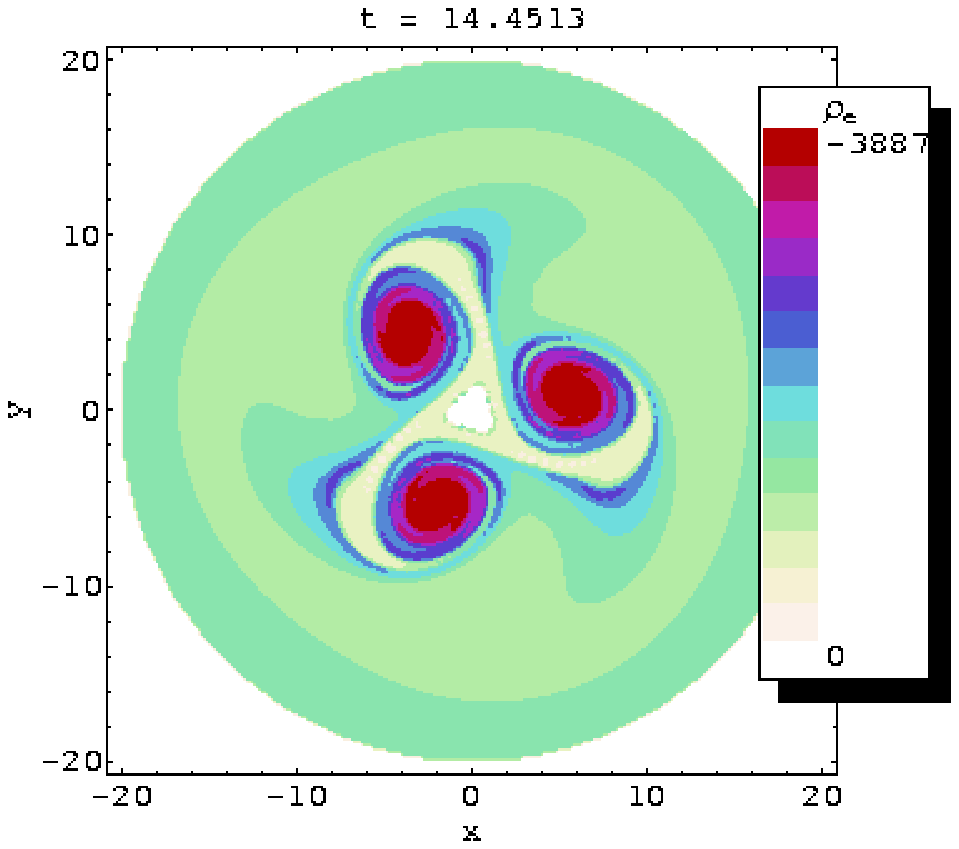} 
  \end{tabular}
  \caption{Snapshot of the charge density corresponding to the runs
    shown in Fig.~\ref{fig:Electrosphere_12} at the transition phase
    between linear and non-linear regime. The $m=3$ patterns are
    clearly identified for sufficiently small times.}
  \label{fig:Electrosphere_12_rho}
\end{figure*}
\begin{figure*}
  \centering
  \begin{tabular}{cc}
    \includegraphics[scale=0.7]{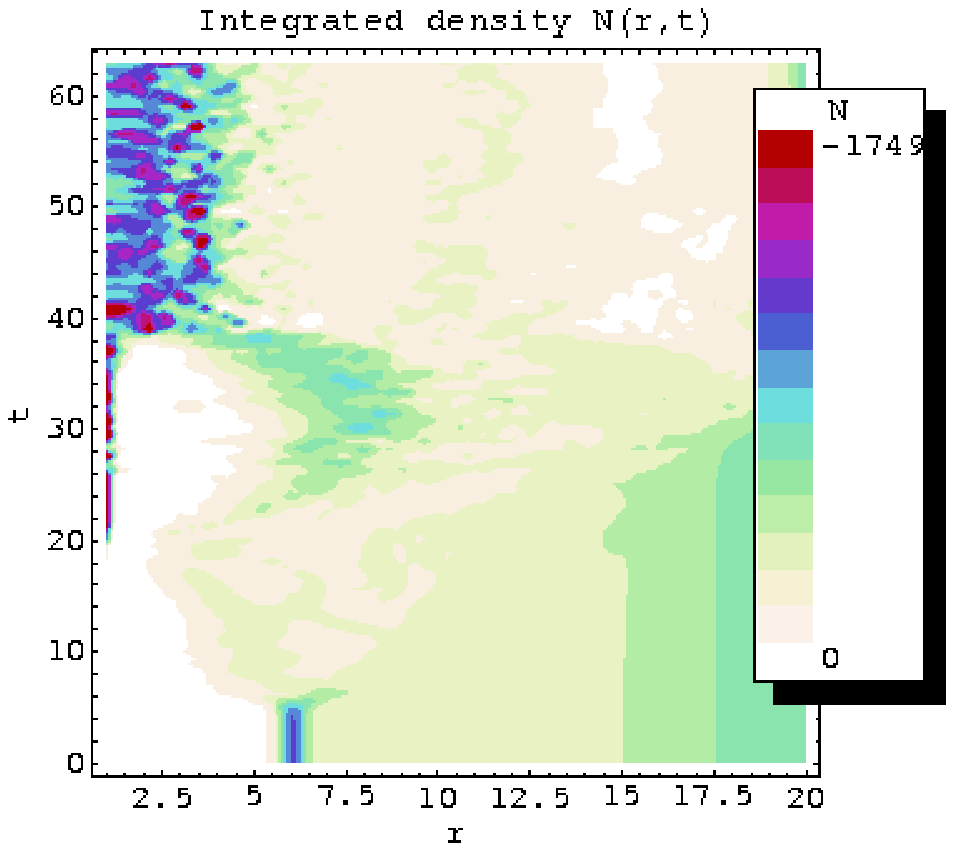} &
    \includegraphics[scale=0.7]{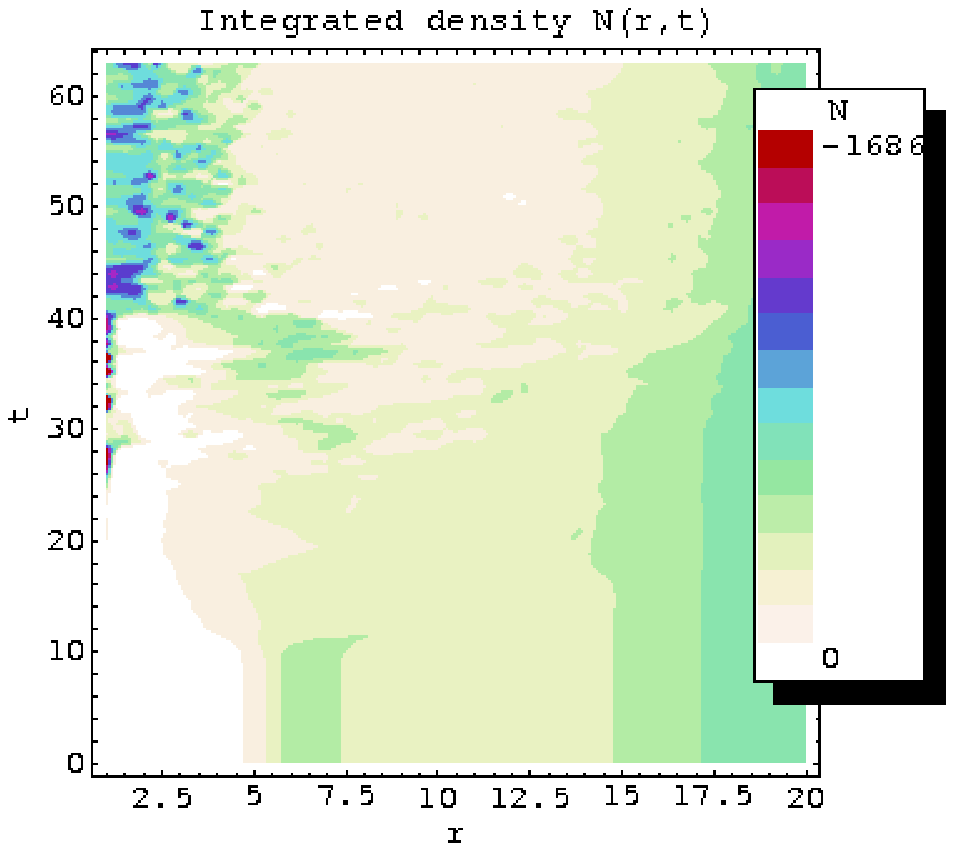} 
  \end{tabular}  
  \caption{Time evolution of the azimuthally integrated charge density
    $\mathcal{N}$ associated with the runs shown in
    Fig.~\ref{fig:Electrosphere_12}.}
  \label{fig:Flux_Densite_Elec_12}
\end{figure*}
In this paper, we show results for the rotation profiles $\Omega_1$
and $\Omega_2$ and a perturbation pattern $m=3$. Note that the plasma
is entirely filling the cylinder from $W_1$ to $W_2$ without vacuum
gaps. First, the evolution of the electrostatic energy is shown in
Fig.~\ref{fig:Electrosphere_12}. In both cases, the observed growth
rates during the linear stage are recovered, in agreement with our
linear analysis (compare the black dots with the red straight line)
for $t<8$ in the first run and $t<12$ in the second case.  After this
first stage, a transition to non-linear regime appears. The density
perturbation becomes significant and is clearly recognizable on the
density map, Fig.~\ref{fig:Electrosphere_12_rho}. Because of small
irregularities, the 3 vortices have different rotational speeds, some
of them overtake the slowest one and coalesce to a larger unique
vortex. Eventually, the last vortex is overtaken and in the final
situation, only one vortex survives and drags towards the neutron star
surface (the inner wall). Indeed, inspecting
Fig.~\ref{fig:Flux_Densite_Elec_12}, the highest charge densities are
found close to $R_1$.  As a conclusion, here again, we found no
significant charge escaping the system and trying to reach the outer
wall.

To summarize these simulations, we demonstrated that our 2D PIC code
is able to accurately reproduce the growth rates of the diocotron
instability in the linear stage of its development in accordance with
our previous analytical study.

The long time evolution in the non-linear stage showed no significant
particle diffusion across magnetic field lines. Current flowing in
the system is not efficient.

Nevertheless, the pulsar magnetosphere is believed to be subject to
copious pair creation feeding the electrosphere with highly
relativistic freshly born electrons and positrons. This external
source of charge can drastically change the long time behavior of the
diocotron instability.  That is why in the next and last paragraph, we
add some particle injection mechanism into the plasma column (or
electrosphere) in order to represent in a simple manner the pair
cascade phenomenon.

\section{RESULTS WITH PARTICLE INJECTION}
\label{sec:ResultsWith}

The most interesting case is certainly the one assuming that the
non-neutral plasma is not isolated but an open system from which
particles can freely enter or exit. Our main focus is on the
electrospheric plasma around a neutron star. Close to its surface, due
to the high magnetic field intensity, some quantum electrodynamical
processes can convert photons into electron-positron pairs. Therefore,
in a crude way, we can think of it as a (time-dependent) source
feeding the electrosphere with charged particles. Moreover, these
charges are responsible for synchrotron and curvature radiation. All
these phenomena can drastically affect the behavior of the diocotron
instability.

Our goal in this last section is to study in detail the
first mentioned effect, namely an external source of charges feeding
the plasma with a flux of particles. In the most general picture, the
spatial and time behavior can be quite complicated. To simplify, we
will assume that the source is located at an arbitrary fixed radius
and that the particle flux is constant in time.


In order to investigate the consequences of particle injection into
the system, we run exactly the same configurations as above but add a
source of particle which are injected at a radius chosen to be
$r=2\,W_1$.  Some simulation results are described in the following
subsections.

\subsection{Plasma column}

First, we consider again the plasma annulus with constant charge
density and trapped between the two conducting walls.  In all these
runs, we adopt exactly the same conditions as before except that we
remove the initial density perturbation. Indeed, we are not interested
in the linear growth rate anymore. In this way, no matter what is the
exact pattern of the perturbation.

\begin{figure*}
  \centering
  \begin{tabular}{cc}
    \includegraphics[scale=0.7]{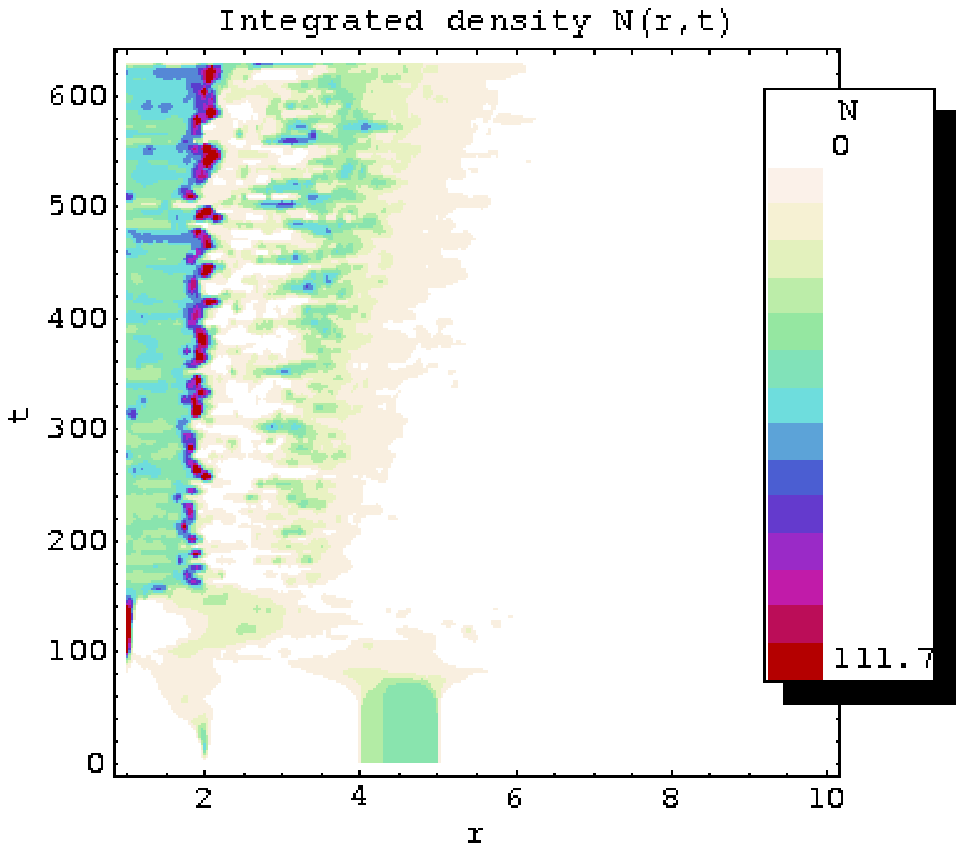} &
    \includegraphics[scale=0.7]{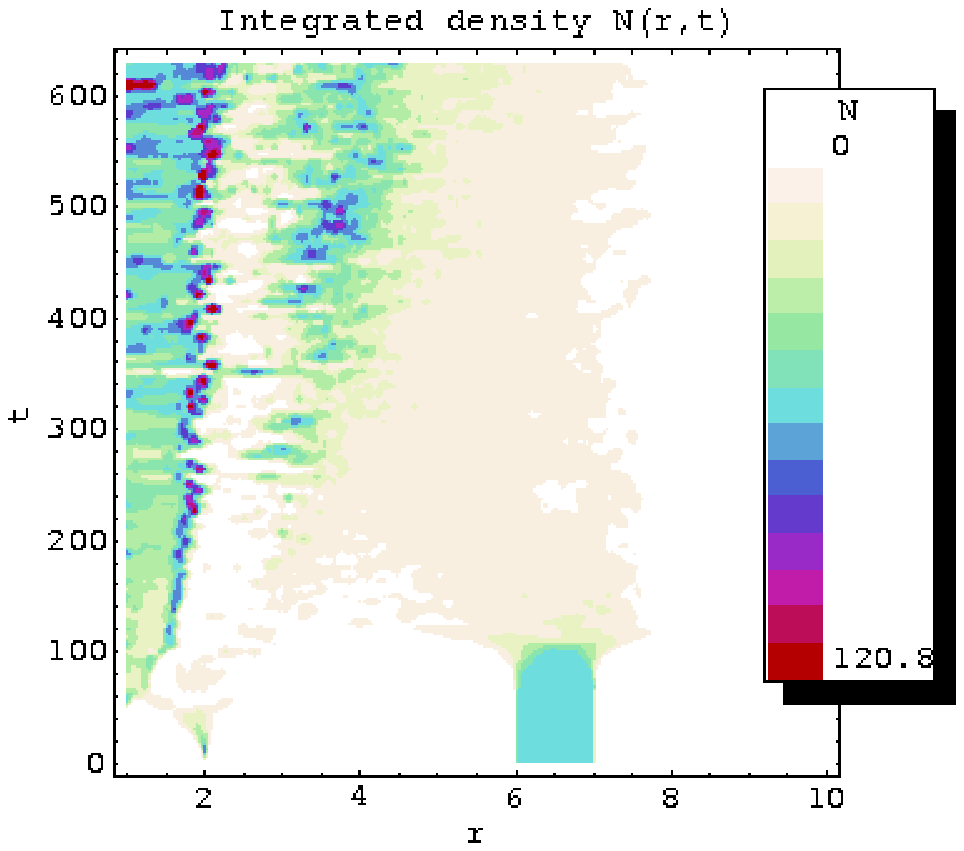} 
  \end{tabular}  
  \caption{Time evolution of the azimuthally integrated charge density
    $\mathcal{N}$ in the plasma column. On the left, a slow increase in
    the particle radial extension is visible, whereas for the case on
    the right, a longer time scale is need to clearly recognize this
    effect.}
  \label{fig:Flux_Densite_Injection_12}
\end{figure*}

\begin{figure*}[htbp]
  \centering
  \begin{tabular}{cc}
    \includegraphics[scale=0.7]{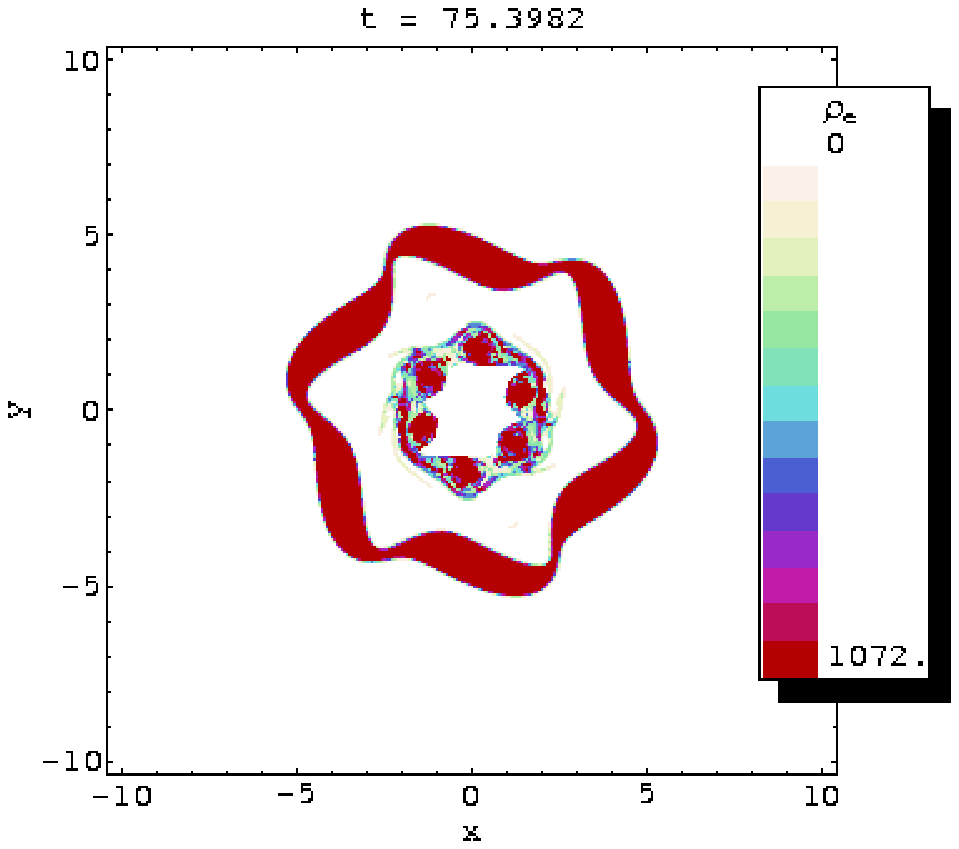} &
    \includegraphics[scale=0.7]{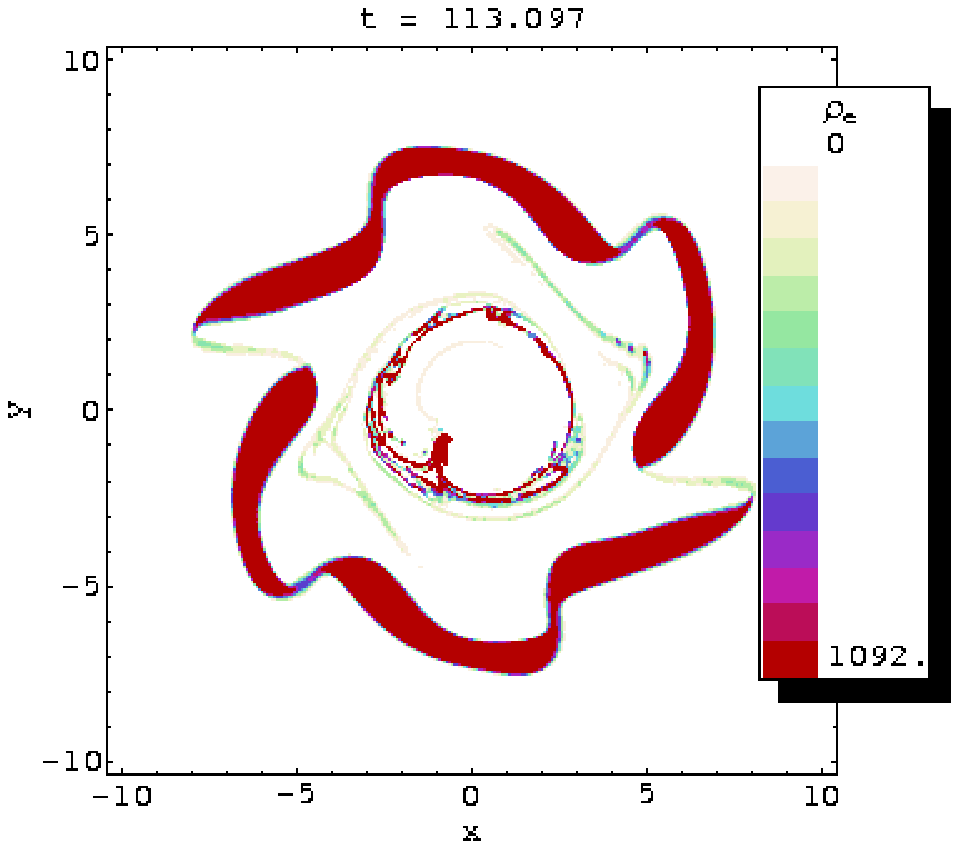} 
  \end{tabular}
  \caption{Snapshot of the charge density in the plasma column having
    the same initial conditions as in Fig.~\ref{fig:Colonne_m37_rho}.
    Due to particle injection, a second inner ring appears, with size
    slowly growing in time. The weak initial particle noise is
    sufficient to ignite the unstable modes which interact with the
    inner annulus.}
  \label{fig:Injection_12_rho}
\end{figure*}

The integrated density is shown in
Fig.~\ref{fig:Flux_Densite_Injection_12} and should be compared with
Fig.~\ref{fig:Flux_Densite_Colonne37}. Note that the color scales are
different in both situations. At the beginning of the runs, the plasma
annulus rotates around its axis without any significant perturbation.
However, as time goes, around $t=60$ and $t=100$ for the first and
second run respectively, the equilibrium configuration has been
destroyed and particles shrink down to the inner wall. As the
simulations start, the location of the source is easily determined on
these maps. Particles are injected at a radius $r=2\,W_1$, depicted on
the plots by a thin strip.
Typical perturbations of the density within the annulus are shown in
Fig.~\ref{fig:Injection_12_rho}. In both cases, it seems that the
fastest growing mode corresponds to a pattern $m=6$. The source of
particles corresponds to the inner annulus. In the first simulation,
on the left figure, there exists a strong interaction between both
rings because they show exactly the same azimuthal pattern, at least,
at the beginning of the non-linear stage. However, this effect is much
less pronounced in the second run, on the right, due to the larger
separation between both these rings.
From a careful comparison between the cases with and without
injection, we conclude that the radial extension at the end of the
simulation is significantly larger (on averaged) when a source is
present. For instance, for the annulus $A1$, the integrated density
does not extend farther than $r=2\,W_1$ when the source is switch off
whereas its extension reaches $r=5\,W_1$ when charges are injected.
The diffusion process would have been even more drastic if the time
span would have been longer. We will make this statement more clear in
the two last sets of run. The annulus configuration was just a
starting point to relate our numerical results to analytically known
solutions for the growth rates.

\subsection{Electrosphere}

Next, we move on to the behavior of the electrosphere which is
actually the main purpose of this work.  We take the same
configurations as in the previous section and add the source located
at $r=2\,W_1$. However, in order to look for particle diffusion
farther away than the outer wall located at $r=20\,W_1$, we used an
initial box larger with size $W_1=1$ and $W_2=100$. This allows us to
investigate large radial excursion of the particle and strong
indication that charges effectively diffuse across the magnetic field
lines.

\begin{figure*}[htbp]
  \centering
  \begin{tabular}{cc}
    \includegraphics[scale=0.7]{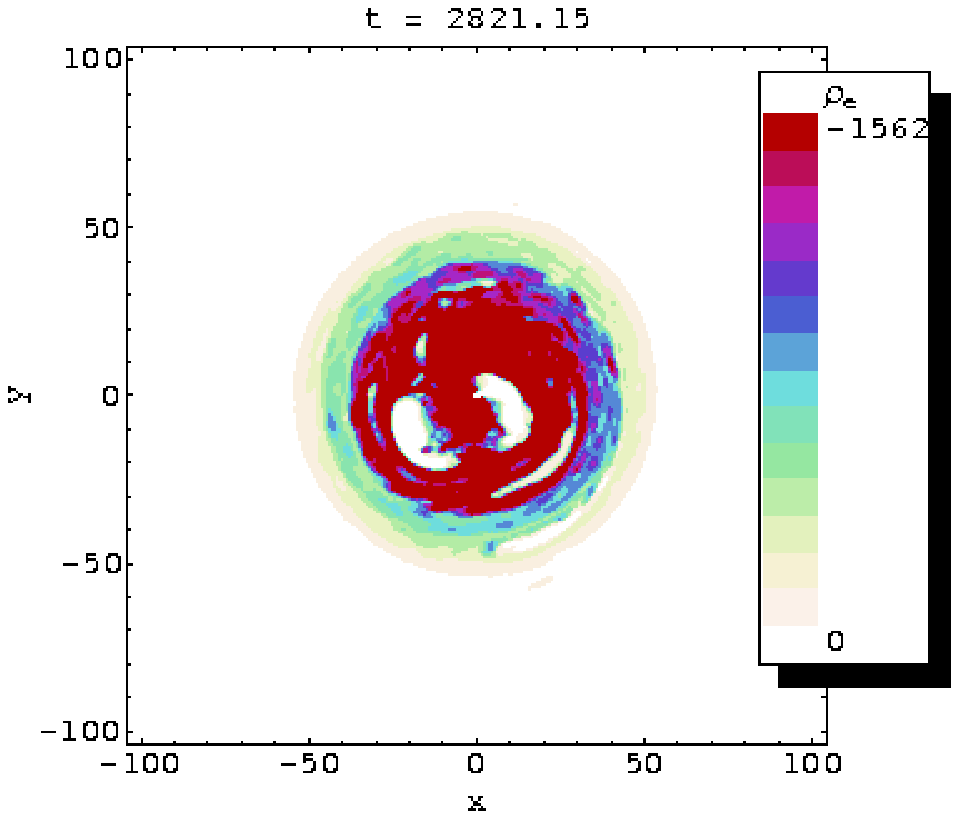} &
    \includegraphics[scale=0.7]{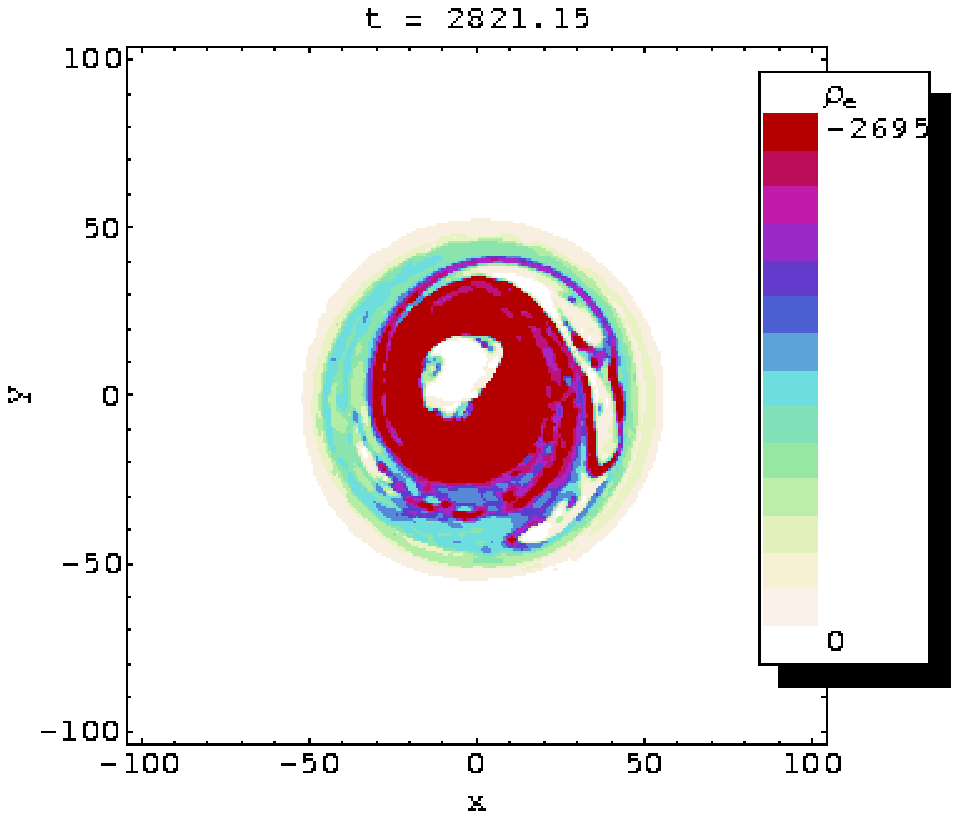} 
  \end{tabular}
  \caption{Snapshot of the charge density for the differentially
    rotating electrosphere, on the left for the case $\Omega_1$ and on
    the right for $\Omega_2$.}
  \label{fig:Injection_Elec_12_rho}
\end{figure*}

\begin{figure*}
  \centering
  \begin{tabular}{cc}
    \includegraphics[scale=0.7]{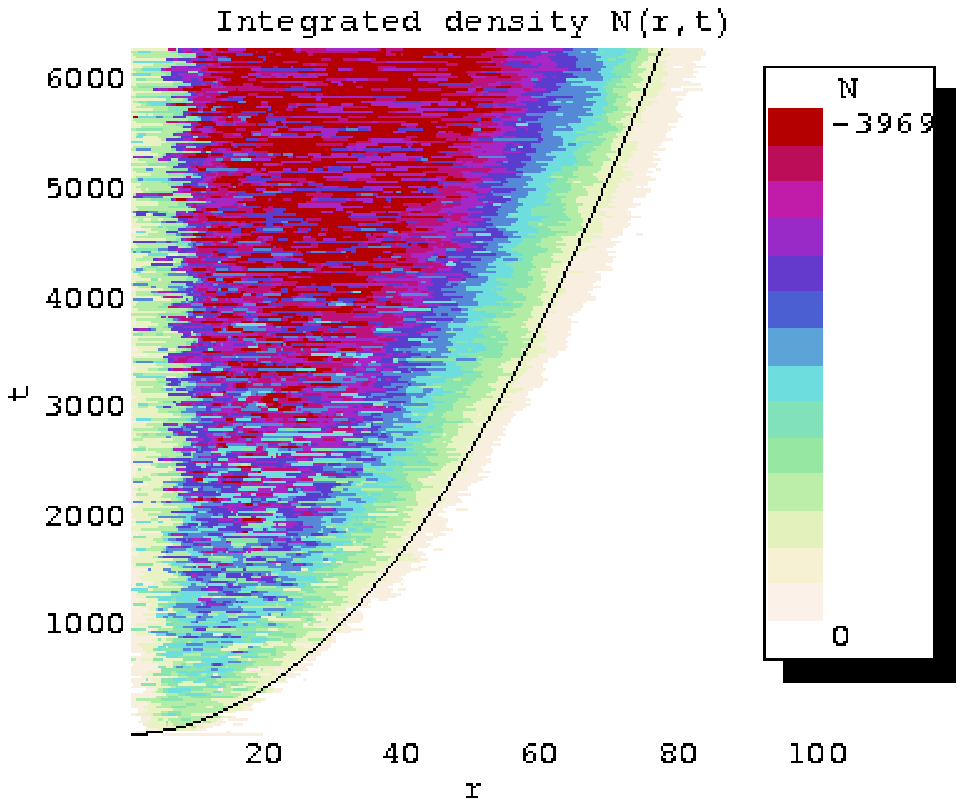} &
    \includegraphics[scale=0.7]{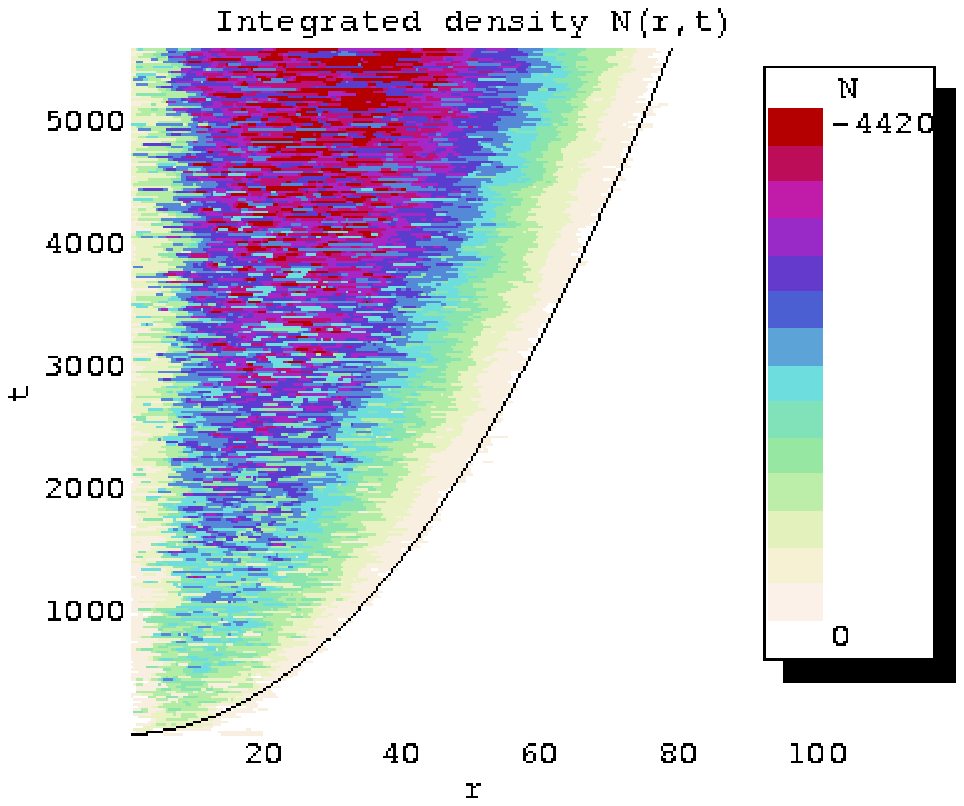} 
  \end{tabular}  
  \caption{Time evolution of the azimuthally integrated charge density
    $\mathcal{N}$ in the electrosphere for the case $\Omega_1$ on the
    left and for $\Omega_2$ on the right.}
  \label{fig:Flux_Densite_Injection_Elec_12}
\end{figure*}

Let us have a look on the charge density at the end of the runs. The
final snapshot is given in Fig.~\ref{fig:Injection_Elec_12_rho} for
$\Omega_1$ on the left, and $\Omega_2$ on the right. We recall that at
the first place, the electrospheric plasma was only filled between
$r=W_1$ and $r=20\,W_1$. In the above pictures, we clearly see a non
negligible charge density up to radii larger than $(40-50)\,W_1$
or so. This is an unambiguous evidence of the efficiency of the
diocotron instability to generate a current across the field lines.
This fact is supported by inspecting
Fig.~\ref{fig:Flux_Densite_Injection_Elec_12}. The integrated density
shows a monotonic and regular expansion. During a first rearrangement
state, the plasma shrink down and approaches the neutron star surface.
This is the meaning of the thin horizontal strip in green-yellow
between $r=W_1$ and $r=20\,W_1$ for time less than
roughly $t<20$. Thus a gap is forming, but after sufficient particles
have been injected into the system, these vacuum regions will be
replenished and expand even farther radially outwards.

\subsection{Vacuum}

Finally, it is also possible to start with the extreme situation containing
no particle, i.e. vacuum. This is certainly the most relevant case to
show the efficiency of particle transport across magnetic field lines.

\begin{figure*}[htbp]
  \centering
  \begin{tabular}{cc}
    \includegraphics[scale=0.7]{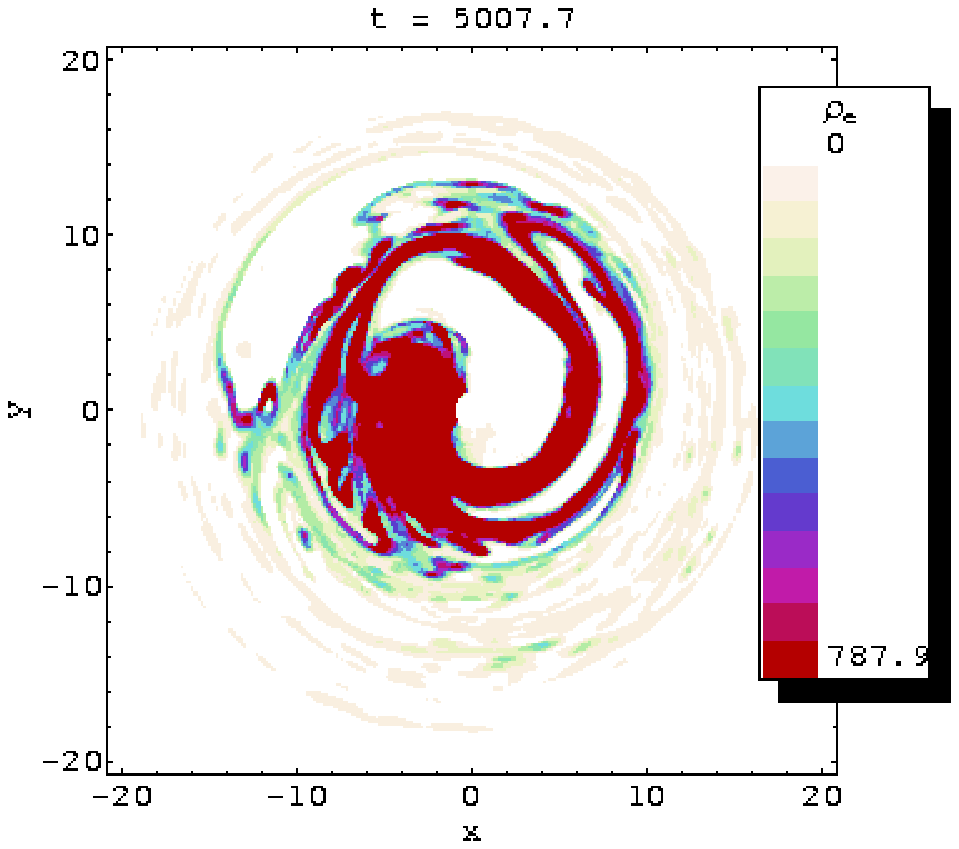} &
    \includegraphics[scale=0.7]{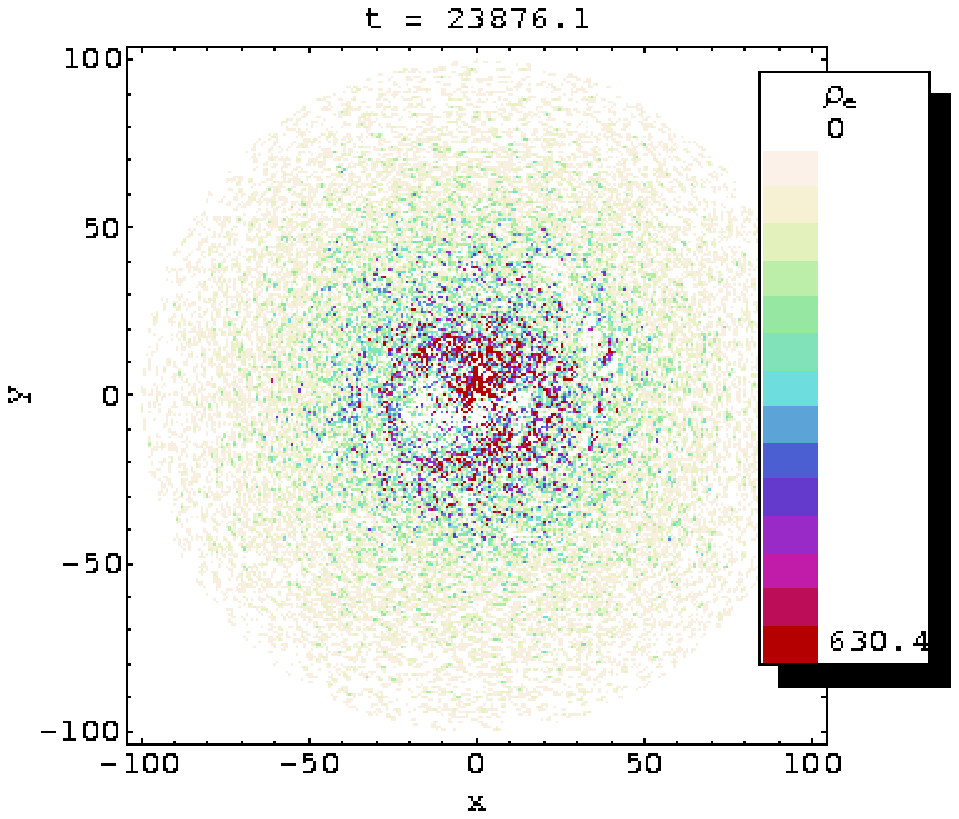}
  \end{tabular}
  \caption{Snapshot of the charge density at the final time of the
    simulation, starting with vacuum.}
  \label{fig:Vide_12_rho}
\end{figure*}

\begin{figure*}
  \centering
  \begin{tabular}{cc}
    \includegraphics[scale=0.7]{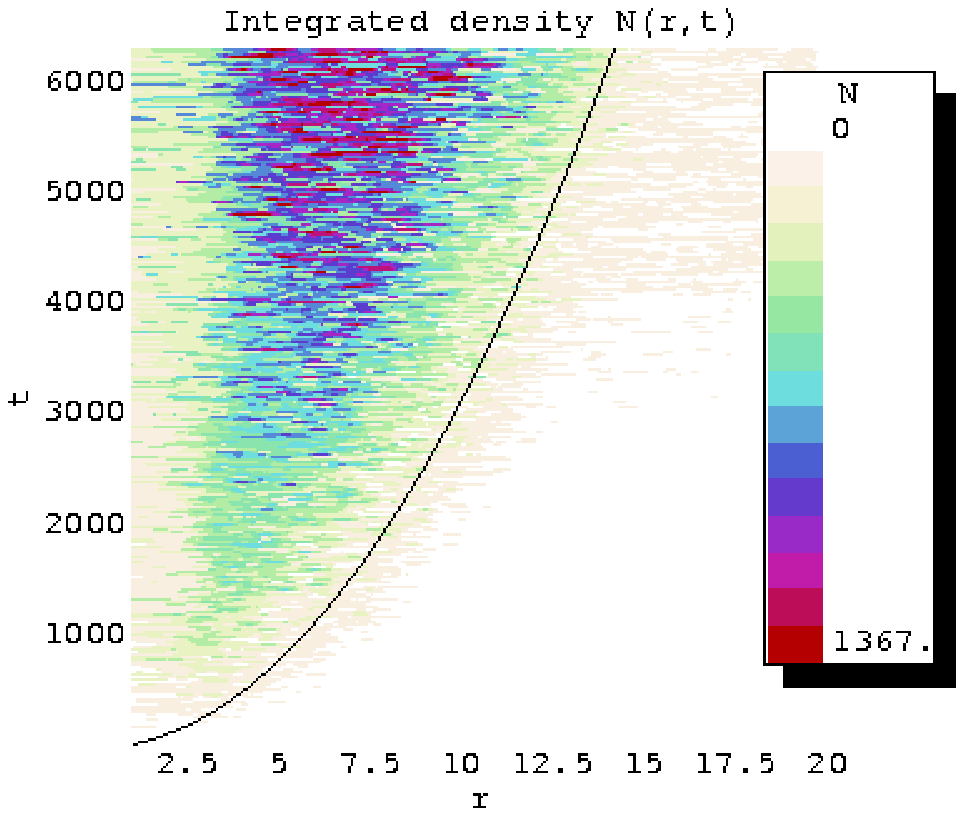} &
    \includegraphics[scale=0.7]{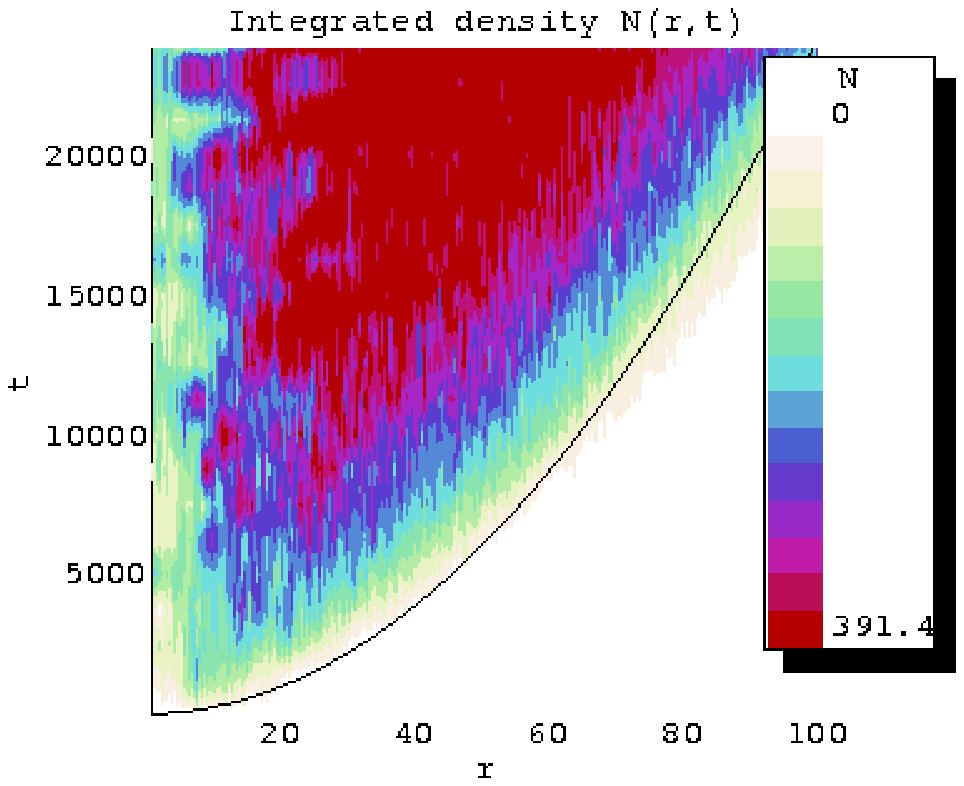} 
  \end{tabular}  
  \caption{Time evolution of the azimuthally integrated charge density
    $\mathcal{N}$, starting with vacuum.}
  \label{fig:Flux_Densite_Vide_12}
\end{figure*}

As in the previous cases, we inject particles at $r=2\,W_1$ and let
the system evolve in a self-consistent manner according to the
electric drift approximation. In the first run, the two conducting
walls are relatively close to each other, with $W_1=1$ and
$W_2=20\,W_1$.  In this way, we are able to see what is happening in
the region close to the injection radius. Depending on the particle
injection flux, the plasma column is growing radially outwards more or
less quickly. As seen in the density snapshot shown on the left part
of Fig.~\ref{fig:Vide_12_rho}, at the end of the simulation, particles
fill almost all the space between the two walls.  Charge transport
across magnetic field lines seems to be very efficient. This is
confirmed by inspecting the left part of
Fig.~\ref{fig:Flux_Densite_Vide_12} in which we see and monotonic
increase in outer edge of the integrated charge density. Starting with
$r$ around $2\,W_1$, the plasma annulus has non negligible density at
$r=13\,W_1$ at the end of the run.

Finally, in order to clearly pick out the long time behavior of this
diocotron instability, we performed a last run with higher resolution
$N_r \times N_\varphi = 400 \times 512$ as well as larger distance
between the conducting walls, $W_1=1$ and $W_2=100$. Therefore we are
in a configuration suitable to investigate the non-linear long time
evolution of the instability. The most relevant results are exposed on
the right part of Fig.~\ref{fig:Vide_12_rho} and
Fig.~\ref{fig:Flux_Densite_Vide_12}, for the density and the
integrated density respectively. Here again, for a sufficient long
time, plasma fills the entire vacuum and spreads out in whole space.
It is clearly seen that the barycenter of the charges drifts to larger
and larger radii. This last run is an unambiguous proof demonstrating
the crucial role of this non-neutral plasma instability for the
modeling of pulsar magnetosphere. Plasma injection in the
electrosphere will unavoidably push particles farther away from the
neutron star.


\subsection{Discussion}
\label{sec:Discussion}

We showed that the non-neutral plasma behaves very differently when
isolated compared to the situation in which a source of particles is
added. In the former case, charges tend to migrate radially inwards
whereas in the latter case, the total charge of the system is not
conserved and they are allowed to move radially outwards. This
migration to larger distances is induced by the increasing electric
repulsion between particles. In the initial (unstable) equilibrium
state, the Lorentz force,~$q\,\vec{E} + \vec{v}\wedge\vec{B}$,
compensates exactly the centrifugal force.  The self-field parameter,
defined by~$s_e = 2\,\omega_p^2/\Omega_c^2 = 2\, n_{\rm e} \, m_{\rm
  e} / B^2$, is such that an equilibrium is permitted. For a constant
density plasma column it corresponds to $s_e \le 1$. We introduced the
plasma frequency (squared) by $\omega_p^2 = n_{\rm e}\,e^2/m_{\rm
  e}\,\varepsilon_0$, the cyclotron frequency by $\Omega_c =
e\,B/m_{\rm e}$, the mass of the particle $m_{\rm e}$ (electron or
positron in our case) and the particle density number $n_{\rm e}$.

Finally, due to the evident dichotomy between an isolated system for
which migration inwards is observed and a system with injection of
particles and a radially outwards expansion of the plasma as
consequences, we could imagine a situation in between such that no
significant radial motion is induced. However, we do not believe that
such a regime is reachable, because a charge density increase will
inevitably lead to a growing self-field parameter~$s_e$ to values
eventually larger than those expected to find an equilibrium solution
(which is unity for the constant density plasma column).  Remember
that the self-field parameter is by definition proportional to the
particle density number for a fixed magnetic field intensity.  For a
plasma column, it is well known that the limit~$s_e=1$ corresponds to
the Brillouin zone and forbids radial confinement above this limit,
\cite{Davidson1990}. For this reason, there will be no particle
injection rate at which both effects annihilate exactly.  The secular
evolution is always particle diffusion across magnetic field lines to
larger distances.

A qualitative and illustrative picture is as follows. For simplicity
we take the example of the plasma column without inner voids, uniform
density and a constant uniform axial magnetic field. The rotation
profile for this column is therefore constant, in other words, the
plasma column is in solid body rotation. Because the plasma is
confined, the self-field parameter should be less than unity. If some
particles were added to the system, keeping the magnetic field
constant, the particle number density~$n_{\rm e}$ increases and so the
self-field parameter with the same proportion.  When the Brillouin
limit is reached, confinement is broken and particles diffuse to
larger radii. A new equilibrium state could be reached in principle
whenever~$s_{\rm e}$ becomes less than unity due to particle dilution
in a greater volume.  Because the timescale for charge diffusion is
much less than the azimuthal rotation period, we can think about a
slow almost adiabatic motion.  The column always readjusts to a new
"quasi"-equilibrium state.  Thus, electrostatic repulsion is at the
heart of the diffusion process.

Finally, we give an estimate for the time evolution of the outer
boundary of the system in the following way. In our simulations, at a
fixed injection rate, the radial excursion slows down because
particles have more space to "dilute" in (area in 2D (or volume in 3D)
proportional to $2\,\pi\,r\,dr(\,dz)$) and therefore decreasing more
efficiently their density number as $r$ grows.  This is indeed
observed in our runs. The outer edge of the plasma filled
region~$R_{\rm edge}$ can roughly be fitted as evolving in time as a
function of $\sqrt{t}$. Indeed, assuming that the number of particles
injected per unit time is $\mathcal{F}$, they should be contained
within the radius $R_{\rm edge}$. Assuming an uniform density within
the column, starting from $r=0$ at $t=0$, we get $\mathcal{F}\,t\,(dz)
\approx n_{\rm e} \, \pi \, R_{\rm edge}^2(\,dz)$ leading to the
expected $\sqrt{t}$ function
\begin{equation}
  \label{eq:Redge}
  R_{\rm edge} \approx \sqrt{\dfrac{\mathcal{F} \, t}{n_{\rm e} \, \pi}}
\end{equation}
This curve is shown as a black solid parabola line on
Fig.~\ref{fig:Flux_Densite_Injection_Elec_12} and
\ref{fig:Flux_Densite_Vide_12}. The constant factor in front of the
$\sqrt{t}$ was adjusted individually for each run.

\section{CONCLUSION}
\label{sec:Conclusion}

We designed a 2D electrostatic PIC code to study the long time
non-linear evolution of the diocotron instability in a background
electromagnetic field with a particular emphasize on the pulsar
magnetosphere where we believe that non-neutral plasmas made of pairs
form an electrosphere known to be diocotron unstable. 

We first check our PIC code by looking for the linear growing stage of
the diocotron unstable modes. We found very good agreement between the
analytical predictions from the linear analysis and the growth rates
derived from our PIC code.

These large growth rates quickly destroy the laminar flow pattern due
to differential rotation. Fluid elements merge together to form some
macroparticles with high density and leaving large vacuum space
between them. Moreover, if the system remains isolated, there is no
significant particle transport across magnetic field lines.
Nevertheless, we found that when adding some external source of
charge, there is a net outflow of charge flowing towards the light
cylinder. This source is believed to come from pair creation in the
innermost part of the electrosphere. Non-neutral plasma instabilities
like the one presented here are an efficient way to transport
particles from regions close to the neutron star to the region where
the wind is launched. The connection between the wind and the close
magnetosphere still needs to be explain. We think that our work will
be a fruitful alternative to existing models.

In this approach, we neglected the magnetic field perturbation induced
by the plasma. This assumption is true only for plasma flowing far
away from the light cylinder (and inside it). In order to take into
account the magnetic field back reaction caused by the plasma, we need
to solve the full set of Maxwell equations. This work will be
presented in a forthcoming paper and will extend the results already
founded in \cite{2007A&A...469..843P} about some relativistic
stabilization effects on the diocotron instability.


\end{document}